\documentclass[11pt,a4paper]{article}
\usepackage{jheppub}

\usepackage{graphicx}
\usepackage{epstopdf}
\usepackage{dcolumn}
\usepackage{bm}

\usepackage{amsmath}
\usepackage{amsfonts}
\usepackage{simpler-wick}
\usepackage{appendix}

\DeclareGraphicsExtensions{.eps}
\newcommand{\D}{\text{D}}
\newcommand{\la}{\langle}
\newcommand{\ra}{\rangle}
\renewcommand{\d}{\text{d}}

\begin{document}

\title{Single-trace current correlators for 2d models of 4d gluon scattering}

\author{Sean Seet}
\affiliation{%
 School of Mathematics and Maxwell Institute for Mathematical Sciences, \\ University of Edinburgh
}%
\emailAdd{sseet@ed.ac.uk}
\date{\today}

\abstract{Correlators of affine Kac-Moody currents evaluate to a sum over multi-color-trace contributions. We present a `single-trace current system' defined on a genus 0 Riemann surface, whose correlators evaluate to precisely the single-trace term in the evaluation of a standard affine Kac-Moody at level 1. We demonstrate that using the single-trace current system in genus 0 Berkovits-Witten twistor string correlators gives pure $\mathcal{N}  = 4$ super Yang-Mills tree amplitudes, and explain how this works both at the level of the string theory correlator and from the perspective of the effective field theory. We explain how using the single-trace current system in genus 0 correlators of the ambitwistor string gives pure Yang-Mills tree amplitudes, and briefly discuss a proposal to go to higher loops.
}

\maketitle


\section{Introduction}
The Parke-Taylor formula for the tree-level MHV (i.e two negative helicity, $n-2$ positive helicity gluons) scattering amplitude of Yang-Mills was conjectured by Parke and Taylor \cite{Parke:1986gb} in 1986 and proven by Berends and Giele \cite{Berends:1987me} in 1988
\begin{equation}\label{eq: YM tree level conn MHV amplitude}
    M^{\text{tree}}(i^-, j^-, 1^+ \ldots n^+)
    = i(-g)^{n-2}\delta^4\left(\sum_{k=1}^n p_k\right)\langle i j \rangle^4\sum_{\sigma \in S_n\setminus \mathbb{Z}_n} \frac{\text{tr}(T^{A_{\sigma(1)}}\ldots T^{A_{\sigma(n)}})}{\langle \sigma(1) \sigma(2) \rangle \ldots \langle \sigma(n) \sigma(1) \rangle} 
\end{equation}
where we have used the spinor helicity formalism, writing each external null momentum $p_i^{\alpha \dot \alpha}$, ($\alpha = 0,1$, $\dot \alpha = \dot 0,\dot 1$ and indices are raised and lowered by the antisymmetric $\epsilon^{\alpha \beta},\epsilon^{\dot\alpha \dot\beta}$ Levi-Civita symbol) in terms of a dotted and undotted spinor
\begin{equation}
    p_i^{\alpha \dot \alpha} = i^{\alpha}\bar i^{\dot \alpha}, \quad p_i \cdot p_j = \underbrace{i^{\alpha}j_\alpha}_{=: \la i j \ra} \,\,\times \,\,\underbrace{\bar i^{\dot \alpha} \bar j_{\dot \alpha}}_{=:[ij]} .
\end{equation}
The sum over $\sigma$ runs over $S_n \setminus \mathbb{Z}_n$, the set of all permutations of $n$ elements with the cyclic permutations removed. Each $T^{A_i}$ is an adjoint generator of the Yang-Mills gauge group, encoding the choice of generator associated to the $i$th gluon wavefunction used in the scattering process. The denominator in the sum is the product $\prod_{i=1}^n \la \sigma(i) \sigma(i+1)\ra$, where $(\sigma(n+1))^\alpha := (\sigma(1))^\alpha $.

The fact that the profusion of gluon Feynman diagrams sums to this remarkably simple formula suggests a hidden underlying structure. The first step towards uncovering this structure was Nair's observation in 1988 \cite{Nair:1988bq} that the sum $\sum \frac{\text{tr}(T^{A_{\sigma(1)}}\ldots T^{A_{\sigma(n)}})}{\langle \sigma(1) \sigma(2) \rangle \ldots \langle \sigma(n) \sigma(1) \rangle}$ could be interpreted in terms of the evaluation of an affine Kac-Moody current correlator at level $k$, defined on $\mathbb{CP}^1$: 
\begin{multline}
    \la \prod_{i=1}^n j^{A_i}(i^\alpha) \ra = k\underbrace{\sum_{\sigma \in S_n\setminus \mathbb{Z}_n} \frac{\text{tr}(T^{A_{\sigma(1)}}\ldots T^{A_{\sigma(n)}})}{\langle \sigma(1) \sigma(2) \rangle \ldots \langle \sigma(n) \sigma(1) \rangle}}_{\text{single-trace}} \prod_{i=1}^n \D i
    \\+ \frac{k^2}{2}\underbrace{\left(\sum_{j,\sigma}\frac{\text{tr}(T^{A_{\sigma(1)}}\ldots T^{A_{\sigma(j)}})}{\langle \sigma(1) \sigma(2) \rangle \ldots \langle \sigma(j) \sigma(1) \rangle}\frac{\text{tr}(T^{A_{\sigma(j+1)}}\ldots T^{A_{\sigma(n)}})}{\langle \sigma(j+1) \sigma(j+2) \rangle \ldots \langle \sigma(n) \sigma(j+1) \rangle}\right)}_{\text{double-trace}}\prod_{i=1}^n \D i + \ldots
\end{multline}
In Nair's construction, the $\mathbb{CP}^1$ in question is the projectivisation of the spinors $i^\alpha$ that encode the null momenta of the scattered gluons, or equivalently the projective null cone of the origin in Minkowski space, the so-called Celestial Sphere. Nair presciently postulated that there should be a twistorial interpretation of the fact that this 2d current correlator computation was producing a result about 4d gluon scattering, but did not propose a mechanism with which to eliminate the multi-trace terms\footnote{to be precise, he did not address/may not have noticed the appearance of the multi-trace terms} or a mechanism to complete the amplitude by filling in the missing kinematical prefactors. There the matter rested until a series of remarkable papers by Witten and then Berkovits in 2004 \cite{Witten_2004,Berkovits_2004,Berkovits_2004confsugra}.

What has now become known as the Berkovits-Witten twistor string \cite{Berkovits_2004} furnished the mechanism with which to recover the missing kinematical factors. In the evaluation of a genus 0 string correlator of the Berkovits-Witten twistor string, we evaluate an affine Kac-Moody current correlator \textit{a la} Nair, but then integrate over the moduli space of holomorphic degree $d$ maps from the $\mathbb{CP}^1$ into $\mathbb{PT}$, the $\mathcal{N}=4$ supertwistor space.
\begin{align}
    Z^\mathcal{A}(\sigma^\alpha): \quad \mathbb{CP}^1_{\sigma} &\rightarrow \mathbb{PT}
    \\ \sigma^\alpha & \mapsto\sum_{i=0}^d Z_i^{\mathcal{A}} (\sigma^1)^{d-i}(\sigma^0)^{i}
\end{align}
and sum over all $d \in \mathbb{Z}^+$, giving the string correlator
\begin{equation}
    \la \text{String correlator} \ra = \sum_{d=1}^n g^{2d} \oint \underbrace{\frac{\bigwedge_{i=0}^d\d^{4|4} Z_i}{\text{Vol}(GL(2,\mathbb{C})}}_{\text{int. over deg. $d$ maps } Z^{\mathcal{A}}(\sigma)} \int_{(\mathbb{CP}^1)^n}\left\la\prod_{j=1}^n  j^{A_j}(\sigma_j)\right\ra \underbrace{\prod_{j=1}^n f_j(Z(\sigma_j))}_{\text{ext. wavefns.}},
\end{equation}
where the $f_j(Z)$ are external twistor wavefunctions for the $\mathcal{N}=4$ gluon multiplet.

At order $k^1$, i.e picking out just the single-trace part of the current correlator, this string correlator gives the RSVW formula for a tree-level connected scattering amplitude of $\mathcal{N}  = 4$ super Yang-Mills \cite{Witten_2004,Roiban_2004}. If the external scattering states are chosen to only be gluons, the $\mathcal{N}  = 4$ super Yang-Mills tree amplitudes coincide with the $\mathcal{N}  = 0$ Yang-Mills amplitude, so that the RSVW formula contains the Parke-Taylor formula as the special case of maps of degree $d=1$ and scattering only external gluons. 

From the evaluation of their string correlators, Witten and Berkovits revealed that the Parke-Taylor formula can be interpreted as arising from manually selecting out the single-trace contribution of current correlators living on holomorphic lines in twistor space, and then integrating over the space of all such lines. The final sticking point is the presence of the multi-trace contributions of the current correlator, which have no interpretation in terms of tree-level Yang-Mills scattering processes and should be interpreted as coming from tree-level exchange of conformal supergravitons \cite{Berkovits_2004confsugra}.

The main result of this paper is the construction of a so-called \emph{single-trace current system}, whose correlators evaluate to only the single trace terms of a conventional affine Kac-Moody algebra
\begin{equation}
    \la nJ^{A_1}(1^\alpha)\prod_{i=2}^n \tilde J^{A_i}(i^\alpha) \ra = \sum_{\sigma \in S_n\setminus \mathbb{Z}_n} \frac{\text{tr}(T^{A_{\sigma(1)}}\ldots T^{A_{\sigma(n)}})}{\langle \sigma(1) \sigma(2) \rangle \ldots \langle \sigma(n) \sigma(1) \rangle} \prod_{i=1}^n \D i
\end{equation}
Use of this current system\footnote{The apparent asymmetry in treatment of the current $J^{A_1}$ and the other current insertions arises as a result of vertex operator descent, from the insertion of $n-1$ picture-changing operators at the locations of $n-1$ of the currents. } in place of the standard affine Kac-Moody in the Berkovits-Witten twistor string means that the genus 0 string correlators evaluate precisely to the RSVW formula, with no terms ignored or removed by hand. This marks the completion of the program begun by Nair's paper in 1988, giving a 2d model whose correlators evaluate precisely to tree level (super) Yang-Mills amplitudes.

Other worldsheet models that are designed to compute tree-level Yang-Mills processes from genus 0 worldsheets include the Ambitwistor string \cite{Mason_2014,Berkovits_2014,siegel2004untwistingtwistorsuperstring,Adamo:2014wea} and the Celestial Holography program \cite{Magnea_2021,Bu:2023vjt,Melton:2024akx,Seet:2024vmh}. The use of a conventional affine Kac-Moody current algebra in ambitwistor strings or \textit{ad-hoc} constructions of putative Celestial duals such as \cite{Melton:2024akx} results in multi-trace processes marking the exchange of color singlets. Replacing the conventional current algebra with our single-trace current algebra removes these multi-trace processes, giving pure tree-level Yang-Mills amplitudes. An interesting implication of our result and those of \cite{Mason_2014,Geyer:2015bja,Roehrig_2018} is that we have a concrete proposal \eqref{eq: 2d theory for yang mills} for a 2d theory (a chiral $\beta \gamma$ system deformed by a nonlocal term) defined on $\mathbb{CP}^1$ that computes perturbative pure Yang-Mills amplitudes correctly at least to 1 loop.

The structure of this paper is as follows. In section \ref{sec: single trace current system}, we describe a free-field realisation of the single-trace current algebra for semi-simple gauge groups $G$ (resp. algebras $\mathfrak{g}$). The key ingredient is realising a level 0 Kac-Moody algebra $\mathfrak{g}_0$ as a subalgebra of the level 1 Kac-Moody superalgebra $\mathfrak{gl}_1(N|N)$ by gauging a system of currents. There is not a canonical way of introducing auxiliary matter with which to construct an anomaly-free gauging, but we discuss two especially simple methods.

In section \ref{sec: application to the BW twistor string}, we discuss the application of the single-trace current algebra to the Berkovits-Witten twistor string. Having done so, we interpret the modifications to the string correlator in terms of the effective field theory and explain the decoupling of conformal supergravity from the computation of tree-level processes with only external gluons. Since multi-trace terms do not contribute to the correlator evaluation, we find that the string correlator now exhibits the correct holomorphic collinear limits and we verify that it correctly reproduces the $\mathcal{N}=4$ super Yang-Mills splitting function. 

In section \ref{sec: Discussion}, we discuss the results and interesting features that merit further investigation. Hyperlinks to important results are provided for a rereader.

\begin{itemize}
    \item Proof that the current algebra evaluation gives a single trace: \ref{subsec: single trace correlator evaluation}
    \item Decoupling of conformal supergravity at genus 0 in the Berkovits-Witten twistor string: \ref{subsec: decoupling of csg}
    \item Collinear limits and splitting function from worldsheet OPEs: \ref{subsec: collinear limits and splitting function}
    \item 2d theory for 4d pure Yang-Mills correct to at least 1-loop: Equation \eqref{eq: 2d theory for yang mills}
\end{itemize}

In Appendix \ref{appendix: symplectomorphism}, we discuss a general trick for generating consistent gaugings in first order theories using the field theory analogue of canonical transformations. The trick is used in the construction of the $\mathfrak{psl}(1|1)$ gauging of subsubsection \ref{subsubsec: psl gaugings}. Since the consistency of the gauging as presented in subsubsection \ref{subsubsec: psl gaugings} can be verified directly, the method of construction is relegated to an Appendix. In Appendix \ref{appendix: BW twistor string} we discuss a minimal set of facts about the effective field theory (EFT) of the Berkovits-Witten twistor string, as a primer for the unfamiliar. An interested reader is directed to \cite{Berkovits_2004confsugra,Mason_2008} for a fuller exposition.
\section{The single-trace current system}\label{sec: single trace current system}
In this section we will discuss the 2d action, gaugings, and vertex operators for the current system that computes the single-trace form factors that appear in a tree-level Yang-Mills process. To set the stage, consider current correlators of a holomorphic Kac-Moody current algebra for a semi-simple Lie group $G$ (Lie algebra $\mathfrak{g}$) at level $k$.
\begin{equation}
    j^A(z_1)j^B(z_2) \sim \frac{k\kappa^{AB}}{z_{12}^2} + \frac{f^{AB}_C j^C(z_2)}{z_{12}}, \quad j^A \in \Omega^{1,0}(\Sigma, \mathfrak{g}),
\end{equation}
where $j^A$ are $(1,0)$-forms valued in the Lie algebra $\mathfrak{g}$, $k$ is the level of the affine Kac-Moody algebra, $\{\kappa^{AB},f^{AB}_C\}$ are the Killing form and structure constants of $\mathfrak{g}$, $z_i$ are coordinates on a $\mathbb{C}$ patch of a Riemann surface $\Sigma$, and $z_{ij}=z_i-z_j$ is convenient shorthand. It was first observed by Nair \cite{Nair:1988bq} that correlators of these currents can generate terms that strongly resemble the Parke-Taylor factor present in the Yang-Mills MHV amplitude
\begin{equation}
    \la \prod j^A(z_i) \ra = \left(-k\sum_{\sigma \in S_n\setminus \mathbb{Z}_n}\frac{\text{tr}(T^{A_{\sigma_1}}\, \ldots \,T^{A_{\sigma_n}})}{z_{\sigma_1\sigma_2}\, \ldots \, z_{\sigma_n\sigma_1}} + \text{multi-trace contributions}\right) \prod_{i=1}^n(\d z^i)
\end{equation}
The first term contains a single color trace, in which $(T^A)^a_b$ is a generator of the adjoint representation of $G$. This factor is recognisably a piece of the tree-level connected MHV scattering amplitude in Yang-Mills, equation \eqref{eq: YM tree level conn MHV amplitude}. However, the evaluation of the current correlator also includes terms at order $k^l$ with $l\in \{2,3,\ldots \lfloor n/2 \rfloor\}$ color traces, which have no interpretation in terms of a tree-level connected scattering process in Yang-Mills.

If we want to extract the single trace part of such a current correlator, a way to operationalise this from the 2d model might be to tune the level $k\rightarrow 0$ in some way so as to suppress the higher trace terms. However, in the realisations of a Kac-Moody from a WZW model or from free fields, the level is an integer and cannot be tuned continuously. It seems that the closest thing we can do is to take the level to be $k=0$ (which is already nontrivial to construct) - but in this case, the single trace term also vanishes.

The proposal of this section is a mechanism by which to make a 2d holomorphic current algebra that is notionally \textit{one step away} from $k=0$ - qualitatively, this is like allowing $k \neq 0$ but $k^2=0$. 

The starting point is the $k=1$ free field realisation of the superalgebra $\mathfrak{gl}_1(N|N,\mathbb{C})$ using bilinears constructed from $N$ free symplectic bosons $\tilde \phi_a, \phi^a$ and $N$ complex fermions $\tilde \rho_a, \rho^a$ on a closed Riemann surface $\Sigma$ of genus $g$. The specific linear combinations $\tilde \phi_a\phi^b + \tilde \rho_a \rho^b$ of the free field bilinears realise the algebra $\mathfrak{gl}_0(N,\mathbb{C})$, from which we can construct the closed subalgebra $\mathfrak{g}_0$ of physical interest by contracting with $(T^A)^a_b$, generators of the adjoint representation of a semi-simple Lie algebra $\mathfrak{g} \subset \mathfrak{sl}(N,\mathbb{C})$ with structure constants $f^{AB}_C$.

To go from $\mathfrak{gl}_1(N|N,\mathbb{C})$ to $\mathfrak{gl}_0(N,\mathbb{C})$, we will gauge two fermionic currents to exclude all of the unwanted bilinears of $\mathfrak{gl}_1(N|N,\mathbb{C})$, leaving only the specific linear combinations $\tilde \phi_a\phi^b + \tilde \rho_a \rho^b$ that generate $\mathfrak{gl}_0(N,\mathbb{C})$. There is not a unique way to do this, but we will describe two especially simple ways. Having done so, after a careful analysis of the gauge-fixing and picture-changing we find that in the case where $\Sigma$ is genus 0, current correlators of this system compute precisely the desired single color trace term, with no multi-trace contributions.

\subsection{Free fields for the single trace current algebra}
We introduce the matter for the single trace current algebra. These consist of $N$ boson and $N$ fermion pairs, where $N$ is the dimension of the fundamental representation of the group $G \subset SL(N,\mathbb{C})$ whose algebra $\mathfrak{g}$ we will wish to describe.
\begin{align}
     &S_{\text{ff}} = \int_{\Sigma} \tilde \phi_a \bar\partial \phi^a +  \tilde \rho_a \bar\partial \rho^a, \quad a = 1, \ldots, N\nonumber
     \\
     &\tilde \rho_a \in \Pi\Omega^0(\Sigma, K^{1/2}_{\Sigma}), \quad \, \rho^a \in \Pi\Omega^0(\Sigma, K^{1/2}_{\Sigma}), \nonumber
     \\&\tilde \phi_a \in \Omega^0(\Sigma, K^{1/2}_{\Sigma}), \quad \, \,\,\,\,\phi^a \in \Omega^0(\Sigma, K^{1/2}_{\Sigma})
\end{align}
The free fields obey the fundamental free field OPEs
\begin{align}
    \tilde \phi_a(z_1) \phi^b(z_2) &\sim \frac{-\delta^b_a}{z_1-z_2}  \sim -\phi^b(z_1) \tilde \phi_a(z_2)
    \\
    \tilde \rho_a(z_1) \rho^b(z_2) &\sim \frac{\delta^b_a}{z_1-z_2}  \sim \rho^b(z_1) \tilde \rho_a(z_2)
\end{align}
Note that the sign of the propagator is different for the bosons than for the case of the free fermions. A simple heuristic with which to remember the sign flip in the propagator is the 2 point calculation in the matrix integral
\begin{equation}
    Z = \int \d^n a \d^n b \, e^{aMb} = \begin{cases}
        \frac{1}{\det M} \quad a,b \text{ Grassmann even}
        \\
        \det M \quad a,b \text{ Grassmann odd}
    \end{cases}
\end{equation}
which implies 
\begin{equation}
    \la a_i b^j \ra = \frac{1}{Z}\frac{\partial Z}{\partial M^i_{\,\,j}} = \begin{cases}
        -(M^{-1})_i^{\,\,j} \quad a,b \text{ Grassmann even}
        \\
        (M^{-1})_i^{\,\,j} \quad a,b \text{ Grassmann odd}
    \end{cases}
\end{equation}
so the sign of the propagator is sensitive to the Grassmann parity. Bilinears formed from one tilded field and one untilded field generate the algebra $\mathfrak{gl}_1(N|N,\mathbb{C})$. For $\mathcal{T}_i \in \mathfrak{gl}(N|N,\mathbb{C})$, we have
\begin{align}
J_{\mathcal{T}}&:=
    \begin{pmatrix}
        \tilde \phi_a & \tilde \rho_b
    \end{pmatrix} \underbrace{\begin{pmatrix}
        A^a_c & B^a_d \\ C^b_c & D^b_d
    \end{pmatrix}}_{\mathcal{T}}\begin{pmatrix}
        \phi^c \\ \rho^d
    \end{pmatrix},
\end{align}
Repackaging the free field OPEs to aid computation,
\begin{equation}
    \begin{pmatrix}
        \phi^c(z_1) \\ \rho^d(z_1) 
    \end{pmatrix}\begin{pmatrix}
        \tilde \phi_a(z_2) & \tilde \rho_b(z_2)
    \end{pmatrix} \sim \frac{\begin{pmatrix}
        \delta^c_a & 0_N \\ 0_N & \delta^b_d
    \end{pmatrix}}{z_{12}}, \quad \begin{pmatrix}
        \tilde \phi_a(z_1) & \tilde \rho_b(z_1)
    \end{pmatrix}\begin{pmatrix}
        \phi^c(z_2) \\ \rho^d(z_2) 
    \end{pmatrix} \sim \frac{\begin{pmatrix}
        -\delta^c_a & 0_N \\ 0_N & \delta^b_d
    \end{pmatrix}}{z_{12}}
\end{equation}
the $JJ$ OPE is swiftly evaluated as
\begin{equation}\label{eq: supergroup current relations}
    J_{\mathcal{T}_1}(z_1)J_{\mathcal{T}_2}(z_2) \sim \frac{-\text{str}(\mathcal{T}_1\mathcal{T}_2)}{z_{12}^2}+\frac{J_{[\mathcal{T}_1,\mathcal{T}_2]}(z_2)}{z_{12}},
\end{equation}
where the \textit{supertrace} is defined to be $\text{str}(\mathcal{T}):=\text{tr}(A)-\text{tr(D)}$\footnote{It may seem more natural to say that the level is therefore $-1$. This is due to a clash of conventions: an $N|N$-index conventionally leads on the bosonic indices, so we are expected to write our fields as $(\tilde \phi, \tilde \rho), (\phi, \rho)$. If we had written them with the fermion leading, as is more natural from the field theory point of view, the resulting double pole would just be the supertrace with a $+$ sign. This is the perspective we take when we say that the level is 1.}.

Rather than being interested in $\mathfrak{gl}_1(N|N,\mathbb{C})$, what we really want is $\mathfrak{gl}_0(N,\mathbb{C})$, which is generated by the specific linear combination of bilinears $\tilde \phi_a\phi^b + \tilde \rho_a \rho^b$. The reason we are interested in $\mathfrak{gl}_0(N,\mathbb{C})$ is that the physical algebra of interest, $\mathfrak{g}_0$, is a subalgebra of $\mathfrak{gl}_0(N,\mathbb{C})$ and can be realised by taking particular linear combinations of the $\mathfrak{gl}_0(N,\mathbb{C})$ currents $\tilde \phi_a\phi^b + \tilde \rho_a \rho^b$, as done by, for instance, Wakimoto \cite{Wakimoto:1986gf}. Define the generators $(T^A)^a_b$ of the adjoint representation of $\mathfrak{g}$, with structure constants $f^{AB}_C$
\begin{equation}
    [T^A,T^B] = f^{AB}_C T^C
\end{equation}
where matrix multiplication in our conventions is $(AB)^a_b := A^a_c B^c_b$. Then the currents that realise $\mathfrak{g}_0$ are
\begin{align}
    &\tilde J^A(z_1) := (T^A)^a_b(\tilde \phi_a\phi^b + \tilde \rho_a \rho^b)
    \\
    &\tilde J^A(z_1) \tilde J^B(z_2) \sim \frac{f^{AB}_C \tilde J^C(z_2)}{z_{12}}
\end{align}
as can be verified from the free-field OPEs. Alternatively, these are the following special case of equation \eqref{eq: supergroup current relations}
\begin{align}
    J_{\text{\tiny$\begin{pmatrix}
        T^{A_1} & 0_N \\ 0_N & T^{A_1}
    \end{pmatrix}$}}(z_1)J_{\text{\tiny{$\begin{pmatrix}
        T^{A_2} & 0_N \\ 0_N & T^{A_2}
    \end{pmatrix}$}}}(z_2) &\sim \frac{-\text{str}\left(\text{\tiny{$\begin{pmatrix}
        T^{A_1} & 0_N \\ 0_N & T^{A_1}
    \end{pmatrix}\begin{pmatrix}
        T^{A_2} & 0_N \\ 0_N & T^{A_2}
    \end{pmatrix}$}}\right)}{z_{12}^2}+\frac{J_{\text{\tiny{$\left[\begin{pmatrix}
        T^{A_1} & 0_N \\ 0_N & T^{A_1}
    \end{pmatrix},\begin{pmatrix}
        T^{A_2} & 0_N \\ 0_N & T^{A_2}
    \end{pmatrix}\right]$}}}(z_2)}{z_{12}} \nonumber 
    \\
    &\sim \frac{J_{\text{\tiny{$\begin{pmatrix}
        [T^{A_1},T^{A_2}] & 0_N \\ 0_N & [T^{A_1},T^{A_2}]
    \end{pmatrix}$}}}(z_2)}{z_{12}} \sim \frac{f^{A_1 A_2}_{C}\tilde J^C}{z_{12}},
\end{align}

In order to operationalize going from $\mathfrak{gl}_1(N|N,\mathbb{C})$ to $\mathfrak{gl}_0(N,\mathbb{C})$, we must gauge a set of currents such that only the $\mathfrak{gl}_0(N,\mathbb{C})$ currents are in the BRST cohomology. Since there is a subalgebra $\mathfrak{sl}(1|1) \otimes \mathfrak{gl}(N) \subset \mathfrak{gl}(N|N)$, it is natural to try gauging some of the currents in the $\mathfrak{sl}(1|1)$, in the hopes that we will land on the $\mathfrak{gl}(N)$ factor. Here, $\mathfrak{sl}(1|1)$ is the Lie superalgebra of dimension $1|2$ (i.e 1 Grassmann even and 2 Grassmann odd generators) with the following defining relations on the basis of generators $\{J| G^\pm\}$,
\begin{equation}
    [J, G^\pm]=0, \quad \{G^+, G^- \} = J
\end{equation}
It is clearest to see the $\{J| G^\pm\} = \mathfrak{sl}(1|1)$ currents in terms of the free fields that realise $\mathfrak{gl}_1(N|N, \mathbb{C})$. 

Consider the rewriting of the free field action into a $1|1$ multiplet transforming in the fundamental under the global action of $\mathcal{M} \in SL(1|1)$
\begin{align}
     &S = \sum_{a=1}^N\int_{\Sigma} \begin{pmatrix}
         \tilde \phi_a & \tilde \rho_a
     \end{pmatrix} \bar \partial \begin{pmatrix}
         \phi^a \\ \rho^a
     \end{pmatrix} = \sum_{a=1}^N\int_{\Sigma} \begin{pmatrix}
         \tilde \phi_a & \tilde \rho_a
     \end{pmatrix} \mathcal{M}^{-1} \bar \partial \left(\mathcal{M}\begin{pmatrix}
         \phi^a \\ \rho^a
     \end{pmatrix}\right), \nonumber
\end{align}
Using the Noether procedure, we see that the classical currents associated to each generator of $\mathfrak{sl}(1|1)$ are
\begin{equation}
    J : \tilde \phi_a \phi^a+\tilde \rho_a \rho^a, \quad G^+ : \tilde \phi_a \rho^a, \quad G^- : \phi^a \tilde \rho_a
\end{equation}
Single contractions between these currents realise the commutation relations of $\mathfrak{sl}(1|1)$:
\begin{align}
    (\tilde \phi_a \phi^a+\tilde \rho_a \rho^a)(z_1)(\tilde \phi_a \phi^a+\tilde \rho_a \rho^a)(z_2) &\sim 0
    \\
    (\tilde \phi_a \phi^a+\tilde \rho_a \rho^a)(z_1)(\tilde \phi_a \rho^a)(z_2) &\sim 0
    \\
    (\tilde \phi_a \phi^a+\tilde \rho_a \rho^a)(z_1)(\phi^a \tilde \rho_a)(z_2) &\sim 0
    \\(\tilde \phi_a \rho^a)(z_1)(\phi^a \tilde \rho_a)(z_2) &\sim \frac{-N}{z_{12}^2} + \frac{(\tilde \phi_a \phi^a+\tilde \rho_a \rho^a)(z_2)}{z_{12}} \label{eq: purely free field current OPEs}
\end{align}

The transformation laws of the free fields under $J,G^\pm$ are
\begin{equation}
    \delta_{J}\begin{pmatrix}
        \tilde \phi_a \\ \tilde \rho_a \\\phi^b \\ \rho^b
    \end{pmatrix} = \begin{pmatrix}
        \tilde \phi_a \\ \tilde \rho_a \\-\phi^b \\ -\rho^b
    \end{pmatrix}, \quad \delta_{G^+}\begin{pmatrix}
        \tilde \phi_a \\ \tilde \rho_a \\\phi^b \\ \rho^b
    \end{pmatrix} = \begin{pmatrix}
        0 \\ \tilde \phi_a \\ -\rho^b \\ 0
    \end{pmatrix}, \quad \delta_{G^-}\begin{pmatrix}
        \tilde \phi_a \\ \tilde \rho_a \\\phi^b \\ \rho^b
    \end{pmatrix} = \begin{pmatrix}
        \tilde\rho_a \\ 0 \\ 0 \\ \phi^b
    \end{pmatrix}
\end{equation}
Demanding invariance under $\delta_J$ means that bilinears of the free fields must contain one tilded and one untilded field, leaving all the generators of $\mathfrak{gl}_1(N|N, \mathbb{C})$. Demanding invariance under $\delta_{G^\pm}$ cuts all the various linear combinations of the bilinears that make up $\mathfrak{gl}_1(N|N, \mathbb{C})$ down to just $\tilde \phi_a\phi^b + \tilde \rho_a \rho^b$, the generators of $\mathfrak{gl}_0(N, \mathbb{C})$. In order to see this, recall that all the $\delta_J$-invariant bilinears can be written as bilinear forms,
\begin{equation}
    J_{\mathcal{T}}=\begin{pmatrix}
        \tilde \phi_a & \tilde \rho_b
    \end{pmatrix}\underbrace{\begin{pmatrix}
        A^a_c & B^b_c \\ C^a_d & D^b_d
    \end{pmatrix}}_{\mathcal{T}}\begin{pmatrix}
        \phi^c \\ \rho^d
    \end{pmatrix},
\end{equation}
where $B,C$ are Grassmann-odd and $A,D$ are Grassmann-even $N \times N$ matrices. In this notation, we have
\begin{equation}\label{eq: bilinear form representation of Gpm}
    G^+ = \begin{pmatrix}
        \tilde \phi_a & \tilde \rho_b
    \end{pmatrix}\begin{pmatrix}
        0^a_c & \delta^b_c \\ 0^a_d & 0^b_d
    \end{pmatrix}\begin{pmatrix}
        \phi^c \\ \rho^d
    \end{pmatrix}, \quad G^- = \begin{pmatrix}
        \tilde \phi_a & \tilde \rho_b
    \end{pmatrix}\begin{pmatrix}
        0^a_c & 0^b_c \\ \delta^a_d & 0^b_d
    \end{pmatrix}\begin{pmatrix}
        \phi^c \\ \rho^d
    \end{pmatrix}
\end{equation}
In the usual way, $\delta_{G^{\pm}}J_{\mathcal{T}}$ are computed by taking the $G^{\pm}J_{\mathcal{T}}$ OPEs. With the aid of equation \eqref{eq: supergroup current relations}, the requirement that $\delta_{G^{\pm}}J_{\mathcal{T}} = 0$ states that $B=0=C$ and $A=D$. Therefore, demanding invariance under the action of $G^\pm$ gives us the constraint that brings us from $\mathfrak{gl}_1(N|N, \mathbb{C})$ to $\mathfrak{g}_0(N,\mathbb{C})$.

With this in mind, the strategy is clear: we must construct a BRST operator that imposes the condition that the BRST cohomology is invariant under the action of $J, G^\pm$ as given above. It is not as easy as simply gauging these 3 currents, due to the presence of anomalies. There is no clearly superior method of doing these gaugings, but we will describe two methods which are each `minimal' in some sense, either of which will be satisfactory for our purposes.

\subsection{BRST operator, vertex operators, picture changing}
The obstruction to naively gauging the three currents $J$, $G^+$, $G^-$ is the double pole in equation \eqref{eq: purely free field current OPEs}. In this section, we will describe how to perform the required gaugings and construct a nilpotent BRST operator in two ways. We will discuss the vertex operators and picture changing procedure, which is identical in both of the cases except for some minor details.
\subsubsection{$\mathfrak{psl}(1|1)$}\label{subsubsec: psl gaugings}
If we supplement the existing free field theory with auxillary matter in the form of a free fermion pair
\begin{equation}
    S_{\text{aux}} = \int_\Sigma u \bar \partial v, \quad u \in \Pi\Omega^0(\Sigma,K_\Sigma), \quad v \in\Pi\Omega^0(\Sigma)
\end{equation}
it is possible to construct a realisation of $\mathfrak{psl}(1|1) \cong \mathbb{C}^{0|2}$ that can be consistently gauged. The improved currents to gauge are the following two fermionic currents
\begin{align}
    \Pi^+&:\tilde \phi_a\rho^a +  u
    \\
    \Pi^-&:\tilde \rho_a\phi^a- v(\tilde\phi_a \phi^a+\tilde \rho_a \rho^a)+N \partial v
\end{align}
where we recall that $N=\delta^a_a$. Note that this current system was not an ad-hoc construction, but is a simple example of a more general symplectomorphism trick (Appendix \ref{appendix: symplectomorphism}) for deriving anomaly-free gaugings. It can be checked that the OPE between $\Pi^\pm$ is regular. 
\begin{align}
    \Pi^+(z_1)\Pi^+(z_2) &\sim 0
    \\
    \Pi^+(z_1)\Pi^-(z_2) &\sim (\tilde \phi_a\rho^a + u)(z_1)(\tilde \rho_a\phi^a-v(\tilde\phi_a \phi^a+\tilde \rho_a \rho^a)+N \partial v)(z_2) \nonumber
    \\
    &\sim\underbrace{\frac{- N}{z_{12}^2} + \frac{(\tilde \phi_a \phi^a+\tilde \rho_a \rho^a)(z_2)}{z_{12}}}_{\tilde \phi_a\rho^a,\tilde \rho_a\phi^a \text{ contractions}} -\underbrace{\frac{(\tilde \phi_a \phi^a+\tilde \rho_a \rho^a)(z_2)}{z_{12}}+\frac{N}{z_{12}^2}}_{u, v  \text{ contractions}}=0
    \\
     \Pi^-(z_1)\Pi^-(z_2) &\sim 0
\end{align}

The full action takes the form
\begin{align}
    S &= \int_{\Sigma} \tilde \phi_a \bar\partial \phi^a +  \tilde \rho_a \bar\partial \rho^a + u \bar \partial v \nonumber
    \\ 
    &+\chi_+(\tilde \phi_a\rho^a + u) + \chi_-(\tilde \rho_a\phi^a+v(\tilde\phi_a \phi^a+\tilde \rho_a \rho^a)-N \partial v)
\end{align}
Gauge fixing the gauge fields $\chi_\pm$ to 0 (in the absence of vertex operator insertions with ghost delta functions), the action takes the gauge-fixed form
\begin{align}
    S_{\text{ff }\mathfrak{psl}(1|1)} = \int_{\Sigma} \tilde \phi_a \bar\partial \phi^a +  \tilde \rho_a \bar\partial \rho^a + u \bar \partial v + \beta^+\bar \partial \gamma_+ + \beta^-\bar \partial \gamma_-
    \\
    \beta^\pm \in \Omega^0(\Sigma,K_{\Sigma}), \quad \gamma^\pm \in \Omega^0(\Sigma)
\end{align}
with BRST operator
\begin{equation}\label{eq: psl(1|1) brst operator}
    Q_{BRST}=\oint \gamma_+\Pi^++\gamma_-\Pi^-=\oint \gamma_+(\underbrace{\tilde \phi_a\rho^a}_{G^+} + u) + \gamma_-(\underbrace{\tilde \rho_a\phi^a}_{G^-}-v\underbrace{(\tilde\phi_a \phi^a+\tilde \rho_a \rho^a)}_{J}+N \partial v)
\end{equation}
The nilpotence of $Q_{BRST}$ can be verified by recalling that the $\Pi^\pm$ currents have regular OPEs with each other. From the arguments given below equation \eqref{eq: bilinear form representation of Gpm}, the presence of the $\mathfrak{sl}(1|1)$ currents in $Q_{BRST}$ imply that the only free-field bilinears that are BRST-closed are those of the form $\tilde \phi_a \phi^b + \tilde \rho_a \rho^b$. 
\begin{equation}
    \delta_{Q_{BRST}} \left(\begin{pmatrix}
        \tilde \phi_a & \tilde \rho_b
    \end{pmatrix}\begin{pmatrix}
        A^a_c & B^b_c \\ C^a_d & D^b_d
    \end{pmatrix}\begin{pmatrix}
        \phi^c \\ \rho^d
    \end{pmatrix}\right) = 0 \iff B=0=C \text{ and } A = D
\end{equation}
Neither $v$ nor $u$ are BRST closed, but $\delta(\gamma_+) v$ is. 

We will be interested in computing correlators of the following "fixed" type vertex operator insertion, with ghost delta functions
\begin{equation}    
J^{A}(z):=\left(\delta(\gamma_+)\delta(\gamma_-)\tilde \rho_a(T^A)^a_b \rho^b\right)(z)
\end{equation}
BRST closure of $J^A$ requires the semisimplicity of the gauge group, because the vanishing of the double contractions with the $J$ current requires tracelessness of $T^A$. The use of $n$ such vertex operator insertions means that it is no longer possible to use the residual symmetries of the action to fix the gauge fields $\chi_\pm$ to zero. Rather, $\chi_\pm$ take values in an $n-1+g$ dimensional moduli space. Picking a nice basis of $\Omega^{0,1}$ and integrating out the Grassmann odd moduli \cite{FRIEDAN198693,VERLINDE198795} gives us $n-1+g$ picture changing operators of the form
\begin{equation}\label{eq: picture changing operators}
    \Upsilon^{\pm} := \delta(\beta^\pm) \delta_{Q_{BRST}}\beta^\pm = \delta(\beta^\pm)\Pi^{\pm}.
\end{equation}
Consider the descent of $J^{A}(z)$ with respect to $\Upsilon^+$. In the OPE limit, the ghost delta functions $\delta(\gamma_+)\delta(\beta^+)$ give a simple zero to cancel the simple pole \cite{witten2023superstringperturbationtheoryrevisited} between $\Pi^+=\tilde \phi_a\rho^a + u$ and $J^{A}$, giving
\begin{align}
    &J^{A}(z) = \left(\delta(\gamma_+)\delta(\gamma_-)\tilde \rho_a(T^A)^a_b \rho^b\right)(z)
    \nonumber
    \\
    &\xrightarrow{\text{descends to}} \delta(\gamma^-) (T^{A})^{b}_{c}\tilde \phi_{b} \rho^{c}(z)
\end{align}
Now descending with respect to $\Upsilon^-$, we have
\begin{align}
    &\delta(\gamma^-) (T^{A})^{b}_{c}\tilde \phi_{b} \rho^{c} \nonumber
    \\
    &\xrightarrow{\text{descends to}} (T^{A})^{b}_{c}(\tilde \phi_{b} \phi^{c}+\tilde \rho_{b} \rho^{c})(z) \nonumber 
    \\ &=: \tilde J^{A}(z)
\end{align}
where $\tilde J^A$ is the descended vertex operator, one of the $\mathfrak{g}_0$ currents. 

In analogy with string theory, the descended vertex operator can be added to the action to give us some insight into its role in the theory. 
\begin{equation}
    S[a]:=\int_{\Sigma} \tilde \phi_a (\bar \partial \delta^a_b + (T^C)^{a}_b a_C(z)) \phi^b + \tilde \rho_a (\bar \partial \delta^a_b + (T^C)^{a}_b a_C(z)) \rho^b + \text{aux}
\end{equation}
We see that the integrated vertex operator is the current that couples to an infinitesimal deformation of a background gauge field $a_C$
on the worldsheet for the gauge group $\mathfrak{g}$. That is,
\begin{equation}
    \la \tilde J^{A}(z) \ldots \ra = \int \D(\text{fields}) e^{-S} (\tilde J^{A}(z) \ldots) = \int \D(\text{fields}) \left(-\frac{\delta e^{-S[a]}}{\delta a_{A}(z)}\right)_{a=0} (\ldots)
\end{equation}
Since it is the Noether current that couples to such a gauge field, it is therefore no surprise that $\tilde J^{A}$ acts as a $\mathfrak{g}$ color rotation of the fundamental fields $\tilde \phi_a, \phi^a, \tilde \rho_a, \rho^a$, as this is the Ward identity associated to this current. With this in mind, the vanishing of the single contractions that might have obstructed BRST closure of $\tilde J^{A}$ can be read off immediately from the fact that the $\mathfrak{sl}(1|1)$ currents in the BRST charge are $\mathfrak{g}$ color singlets. In contrast, $J^{A}$ contains the current that couples to a color rotation of only the fermions $\tilde \rho_a, \rho^a$. The presence of the ghost delta functions (and tracelessness of the generators of $\mathfrak{g}$) is what ensures BRST closure of $J^{A}$.

In the case where $\Sigma = \mathbb{CP}^1$, the correlator of $n$ fixed vertex operators is therefore
 \begin{align}\label{eq: psl formalism correlator evaluaton}
     &\left \la v(z_1) \prod_{i=2}^{n} \Upsilon^+(z_i) \prod_{i=2}^{n} \Upsilon^-(z_i) \prod_{i=1}^{n} J^{A_i}(z_i) \right \ra=\left\la v(z_1) J^{A_i}(z_1) \prod_{i=2}^{n} \tilde J^{A_i}(z_i) \right \ra\nonumber
     \\
     &=\left\la \tilde \rho_{a_1} (T^{A_1})^{a_1}_{b_1} \rho^{b_1}(z_1) \prod_{i=2}^{n} \left(\tilde \phi_{a_i} (T^{A_i})^{a_i}_{b_i} \phi^{b_i}(z_i)+\tilde \rho_{a_i} (T^{A_i})^{a_i}_{b_i} \rho^{b_i}(z_i)\right)\right \ra_{\text{free-fields}} \times \nonumber \\
     &\la v(z_1) \delta(\gamma_+(z_1))\delta(\gamma_-(z_1))\ra_{\text{ghosts}} \nonumber
      \\
      &=\left\la \tilde \rho_{a_1} (T^{A_1})^{a_1}_{b_1} \rho^{b_1}(z_1) \prod_{i=2}^{n} \left(\tilde \phi_{a_i} (T^{A_i})^{a_i}_{b_i} \phi^{b_i}(z_i)+\tilde \rho_{a_i} (T^{A_i})^{a_i}_{b_i} \rho^{b_i}(z_i)\right)\right \ra_{\text{free-fields}},
 \end{align}
 where we have chosen the locations of the picture changing operators to coincide with the vertex operators at $z_2, ... z_{n}$, and placed an $v$ field at $z_1$ to saturate the fermionic zero mode. The $v$ must be placed at the position of a $\delta(\gamma_+)$ because it is not BRST invariant on its own.
 
 By integrating out the antighosts $\beta^\pm$ and $u$, we find that $\gamma_{\pm},v$ are holomorphic, and therefore have one zero mode on the Riemann sphere. The explicit insertions of $v, \delta(\gamma_\pm)$ are done against the zero mode integrals, so that the ghost correlator evaluates to 1.
 
 The evaluation of the free-field correlator is deferred to subsection \ref{subsec: single trace correlator evaluation}, where we will see that it evaluates precisely to the desired single-trace term.

\subsubsection{$\mathfrak{sl}(1|1)$}
It may be considered unnatural to introduce an extra free fermion pair $u,v$. It is possible to consistently gauge the currents $J : \tilde \phi_a \phi^a+\tilde \rho_a \rho^a$, $G^+ : \tilde \phi_a \rho^a$, $G^- : \phi^a \tilde \rho_a$ that comprise $\mathfrak{sl}(1|1)$ by doubling the field content in a supersymmetric way. Let us extend the index $a$ to a super-index $\mathcal{A} = a | i$, where $a,i$ run from 1 to $N$, and where the fields with an $a$ subscript (resp. superscript) and $i$ subscript (resp. superscript) have opposite Grassmann parity. The action is therefore
\begin{equation}
    \int_{\Sigma} \tilde \phi_{\mathcal{A}} \bar\partial \phi^{\mathcal{A}} +  \tilde \rho_{\mathcal{A}} \bar\partial \rho^{\mathcal{A}} = \int_{\Sigma} \underbrace{\tilde \phi_{a} \bar\partial \phi^{a} + \tilde \rho_{i} \bar\partial \rho^{i}}_{\text{bosonic}} +\underbrace{\tilde \phi_{i} \bar\partial \phi^{i}+  \tilde \rho_{a} \bar\partial \rho^{a}}_{\text{fermionic}}
\end{equation}
and the currents to gauge are
\begin{align}
     J&=\tilde \phi_{\mathcal{A}} \phi^{\mathcal{A}}+\tilde \rho_{\mathcal{A}} \rho^{\mathcal{A}}
     \\
     G^+&= \tilde \phi_{\mathcal{A}} \rho^{\mathcal{A}} 
     \\
     G^-&=\phi^{\mathcal{A}} \tilde \rho_{\mathcal{A}}
\end{align}
The double pole in the $G^+G^-$ OPE came from the nonvanishing superdimension\footnote{The superdimension of an index $\mathcal{A}$ running over $p$ bosonic and $q$ fermionic values is $p-q$} of the index on the free fields. Now, the index $\mathcal{A}$ runs over $N|N$ values, so the superdimension is 0 and the double pole vanishes
\begin{align}
    G^+(z) G^-(w) &= (\tilde \phi_a \rho^a + \tilde \phi_i \rho^i)(z)(\phi^b\tilde \rho_b + \phi^i \tilde \rho_i)(w)\sim 
    (\tilde \phi_a \rho^a)(z)(\phi^b\tilde \rho_b)(w) + (\tilde \phi_i \rho^i)(z)(\phi^j \tilde \rho_j)(w)\nonumber
    \\
    &\sim \underbrace{\frac{N}{(z-w)^2}+\frac{\tilde \phi_a\phi^a(w) + \tilde \rho_a \rho^a(w)}{z-w}}_{\text{a,b, index contribution}} + \underbrace{\frac{-N}{(z-w)^2}+\frac{\tilde \phi_i\phi^i(w) + \tilde \rho_i \rho^i(w)}{z-w}}_{\text{i,j, index contribution}}\nonumber
    \\ &\sim \frac{J(w)}{z-w}
\end{align}
It can be checked that the $J G^\pm$ OPE is still regular. This means that with this supersymmetrisation of the free fields, the currents $J, G^+,G^-$ can be consistently gauged. Introducing gauge fields for all three of these currents, the action takes the form
\begin{equation}
    \int_{\Sigma} \tilde \phi_{\mathcal{A}} \bar\partial \phi^{\mathcal{A}} +  \tilde \rho_{\mathcal{A}} \bar\partial \rho^{\mathcal{A}} + a_J(\tilde \phi_{\mathcal{A}} \phi^{\mathcal{A}}+\tilde \rho_{\mathcal{A}} \rho^{\mathcal{A}}) + \chi_+ \tilde \phi_{\mathcal{A}} \rho^{\mathcal{A}} + \chi_-\phi^{\mathcal{A}} \tilde \rho_{\mathcal{A}}
\end{equation}
Gauge fixing the gauge fields $a_J, \chi_\pm$ to zero (in the absence of vertex operators with ghost delta functions), we find that the action takes the gauge-fixed form
\begin{align}
    &S=\int_{\Sigma} \tilde \phi_{\mathcal{A}} \bar\partial \phi^{\mathcal{A}} +  \tilde \rho_{\mathcal{A}} \bar\partial \rho^{\mathcal{A}} + \underbrace{\beta^+\bar \partial \gamma_+ + \beta^-\bar \partial \gamma_- + \beta^J \bar \partial \gamma_J}_{S_{\text{ghosts}}} \nonumber
    \\
    &\beta^\pm \in \Omega^0(\Sigma,K_{\Sigma}),\, \gamma_\pm \in \Omega^0(\Sigma),\, \nonumber
    \\
    &\beta^J \in \Pi\Omega^0(\Sigma,K_{\Sigma}),\, \gamma_J \in \Pi\Omega^0(\Sigma),\,
\end{align}
The ghost contributions to the $\mathfrak{sl}(1|1)$ currents are of the usual form $\frac{1}{2}\beta f^{\pm,J} \gamma$, and the BRST operator is
\begin{equation}
    Q_{BRST} = \oint \gamma_+ (\tilde \phi_{\mathcal{A}} \rho^{\mathcal{A}}) + \gamma_-(\phi^{\mathcal{A}}\tilde \rho_{\mathcal{A}}) + \gamma_J(\tilde \phi_{\mathcal{A}} \phi^{\mathcal{A}}+\tilde \rho_{\mathcal{A}} \rho^{\mathcal{A}}) + \gamma_+\gamma_-\beta^J
\end{equation}
Nilpotence can be verified by direct computation. Just as in the section on $\mathfrak{psl}(1|1)$, we will be interested in computing correlators of the following vertex operators with ghost delta functions
\begin{equation}    
J^{A}(z):=\left(\delta(\gamma_+)\delta(\gamma_-)\tilde \rho_a(T^A)^a_b \rho^b\right)(z)
\end{equation}
As before, the evaluation of a correlator requires the use of $n-1+g$ picture changing operators of the form
\begin{equation}
    \Upsilon^{\pm} := \delta(\beta^\pm) \delta_{Q_{BRST}}\beta^\pm
\end{equation}
Consider the descent of $J^{A}(z)$ with respect to $\Upsilon^+$. The ghost delta functions $\delta(\gamma_+)\delta(\beta^+)$ instruct us to pick up the simple pole between $G^{+}=\tilde \phi_{\mathcal{A}}\rho^{\mathcal{A}} + \frac{1}{2}\gamma_{-}\beta^J$ and $J^{A}$, giving
\begin{align}
    &J^{A}(z) = \left(\delta(\gamma_+)\delta(\gamma_-)\tilde \rho_a(T^A)^a_b \rho^b\right)(z)
    \nonumber
    \\
    &\xrightarrow{\text{descends to}} \delta(\gamma^-) (T^{A})^{b}_{c}\tilde \phi_{b} \rho^{c}(z)
\end{align}
Now descending with respect to $\Upsilon^-$, we have
\begin{align}
    &\delta(\gamma^-) (T^{A})^{b}_{c}\tilde \phi_{b} \rho^{c} \nonumber
    \\
    &\xrightarrow{\text{descends to}} (T^{A})^{b}_{c}(\tilde \phi_{b} \phi^{c}+\tilde \rho_{b} \rho^{c})(z) \nonumber 
    \\ &=: \tilde J^{A}(z)
\end{align}
where $\tilde J^A$ is the descended vertex operator. Precisely as before in the section on $\mathfrak{psl}(1|1)$, in the case where $\Sigma = \mathbb{CP}^1$, the correlator of $n$ fixed vertex operators is
 \begin{align}
     &\left \la \gamma_J(z_1) \prod_{i=2}^{n} \Upsilon^+(z_i) \prod_{i=2}^{n} \Upsilon^-(z_i) \prod_{i=1}^{n} J^{A_i}(z_i) \right \ra=\left\la \gamma_J(z_1) J^{A_i}(z_1) \prod_{i=2}^{n} \tilde J^{A_i}(z_i) \right \ra\nonumber
     \\
     &=\left\la \tilde \rho_{a_1} (T^{A_1})^{a_1}_{b_1} \rho^{b_1}(z_1) \prod_{i=2}^{n} \left(\tilde \phi_{a_i} (T^{A_i})^{a_i}_{b_i} \phi^{b_i}(z_i)+\tilde \rho_{a_i} (T^{A_i})^{a_i}_{b_i} \rho^{b_i}(z_i)\right)\right \ra_{\text{free-fields}}\times \nonumber
     \\  &\la \gamma_J(z_1) \delta(\gamma_+(z_1))\delta(\gamma_-(z_1))\ra_{\text{ghosts}}\nonumber
     \\&=\left\la \tilde \rho_{a_1} (T^{A_1})^{a_1}_{b_1} \rho^{b_1}(z_1) \prod_{i=2}^{n} \left(\tilde \phi_{a_i} (T^{A_i})^{a_i}_{b_i} \phi^{b_i}(z_i)+\tilde \rho_{a_i} (T^{A_i})^{a_i}_{b_i} \rho^{b_i}(z_i)\right)\right \ra_{\text{free-fields}}
 \end{align}
 Where we have chosen the locations of the picture changing operators to coincide with the vertex operators at $z_2, ... z_{n}$, and placed a $\gamma_J$ field at $z_1$ to saturate the fermionic zero mode. $\gamma_J$ is not BRST-invariant on its own, and must be co-located with the ghost delta functions. The evaluation of the free-field correlator is deferred to subsection \ref{subsec: single trace correlator evaluation}, where we will see that it evaluates precisely to the desired single-trace term.

 It is important to check that instantonic configurations of $a_J$ do not contribute to the evaluation of the correlator. Summing over instanton number amounts to twisting the tilded and untilded fields by a degree $d$ line bundle and summing over $d \in \mathbb{Z}$
 \begin{equation}
     \tilde{\phi}_{a}, \tilde\rho_i \in \Omega^0(\Sigma, K^{1/2}_{\Sigma} \times \mathcal{L}_d), \quad {\phi}^{a},\rho^i \in \Omega^0(\Sigma, K^{1/2}_{\Sigma} \times {\mathcal{L}_d}^{-1}), \quad
 \end{equation}
 and likewise for the Grassmann-odd fields. 
 \begin{equation}
     \tilde{\phi}_{i}, \tilde\rho_a \in \Pi\Omega^0(\Sigma, K^{1/2}_{\Sigma} \times \mathcal{L}_d), \quad {\phi}^{i},\rho^a \in \Pi\Omega^0(\Sigma, K^{1/2}_{\Sigma} \times {\mathcal{L}_d}^{-1}), \quad
 \end{equation}
 Twisting with $d > 0$ (resp. $d<0$) increases the number of zero modes for the tilded (resp. untilded) fields. Since the fields with an $i$ index decouple from the correlator evaluation, we find that the resulting evaluation is ill-defined due to the divergent bosonic zero-mode integrals over the $\tilde \rho_i$ (resp. untilded) and vanishing fermionic zero-mode integrals over the $\tilde \phi_i$ (resp. untilded). Choosing to regulate the bosonic zero-mode integrals, for instance by making the bosons periodic with a finite volume, we find that the fermionic zero-mode integrals cause the correlator to vanish for all $d \neq 0$.
 
\subsection{Correlator evaluation}\label{subsec: single trace correlator evaluation}
The correlator evaluation is largely identical in both the $\mathfrak{psl}(1|1)$ case and the $\mathfrak{sl}(1|1)$ case. At least for genus 0, there is no clear indicator as to which approach is superior. In the $\mathfrak{psl}(1|1)$ case, we need an additional free fermion system, whereas in the $\mathfrak{sl}(1|1)$ case, we need to double the field content. The approach that is superior (cleaner/more elegant) will depend on the specific application.

In any case, after gauge fixing, picture changing, and integrating out the ghosts and auxillary fields (in the case of $\mathfrak{psl}(1|1)$ it is $u, v$, and in the case of $\mathfrak{sl}(1|1)$, it is $\tilde \phi_i, \phi^i, \tilde \rho_i, \rho^i$), the final step is to evaluate the free-field current correlator
\begin{multline}
    \left\la \tilde \rho_{a_1} (T^{A_1})^{a_1}_{b_1} \rho^{b_1}(z_1) \prod_{i=2}^{n} \left(\tilde \phi_{a_i} (T^{A_i})^{a_i}_{b_i} \phi^{b_i}(z_i)+\tilde \rho_{a_i} (T^{A_i})^{a_i}_{b_i} \rho^{b_i}(z_i)\right)\right \ra_{\text{free-fields}} \\
    = \frac{1}{Z}\int \D(\tilde \phi_a, \phi^a, \tilde \rho_a, \rho^a) e^{S_{\text{ff}}} \left(\rho_{a_1} (T^{A_1})^{a_1}_{b_1} \rho^{b_1}(z_1) \prod_{i=2}^{n} \left(\tilde \phi_{a_i} (T^{A_i})^{a_i}_{b_i} \phi^{b_i}(z_i)+\tilde \rho_{a_i} (T^{A_i})^{a_i}_{b_i} \rho^{b_i}(z_i)\right)\right)
\end{multline}
in which
\begin{align}
     &S_{\text{ff}} = \int_{\Sigma} \tilde \phi_a \bar\partial \phi^a +  \tilde \rho_a \bar\partial \rho^a, \quad a = 1, \ldots, m\nonumber
     \\
     &\tilde \rho_a \in \Pi\Omega^0(\Sigma, K^{1/2}_{\Sigma}), \, \rho^a \in \Pi\Omega^0(\Sigma, K^{1/2}_{\Sigma}), \nonumber
     \\&\tilde \phi_a \in \Omega^0(\Sigma, K^{1/2}_{\Sigma}), \, \,\,\,\,\phi^a \in \Omega^0(\Sigma, K^{1/2}_{\Sigma})
\end{align}
The claim of this section is that in the case where $\Sigma = \mathbb{CP}^1$, the Riemann sphere, this current correlator is a pure single trace contribution
\begin{multline}\label{eq: statement of single trace Parke-Taylor}
    \left\la \tilde \rho_{a_1} (T^{A_1})^{a_1}_{b_1} \rho^{b_1}(z_1) \prod_{i=2}^{n} \left(\tilde \phi_{a_i} (T^{A_i})^{a_i}_{b_i} \phi^{b_i}(z_i)+\tilde \rho_{a_i} (T^{A_i})^{a_i}_{b_i} \rho^{b_i}(z_i)\right)\right \ra_{\text{free-fields on } \mathbb{CP}^1}
     \\= \sum_{\sigma \in S_n\setminus \mathbb{Z}_n} \frac{\text{tr}(T^{A_{\sigma_1}}\, \ldots \,T^{A_{\sigma_n}})}{z_{\sigma_1\sigma_2}\, \ldots \, z_{\sigma_n\sigma_1}} 
\end{multline}
Before we go into the details of the proof, the intuition\footnote{brought to my attention by David Skinner} is the following. Since there are no zero modes for any of the free fields, the only nonzero contributions to the correlator will be from taking all possible contractions until there are no more quantum fields present. The result of the evaluation will therefore be a sum over meromorphic functions of $z_i$ which are completely characterised by their singularities as $z_{ij} \rightarrow 0$ for all possible $i,j$. Since they are characterised by their singularities, each OPE channel therefore contributes a term to the sum over all possible Wick contractions. With this in mind, observe that the OPEs between the fermionic bilinear $j^{A}:=\tilde \rho_{a} (T^{A})^{a}_{b} \rho^{b}$ and the sum of fermionic and bosonic bilinears $j^A + b^A := \tilde \rho_{a} (T^{A})^{a}_{b} \rho^{b} + \tilde \phi_{a} (T^{A})^{a}_{b} \phi^{b}$ takes the form
\begin{align}\label{eq: traces weighted with k}
    j^A(z_1) (j+b)^B(z_2) &\sim \frac{k\kappa^{AB}}{z_{12}^2} + \frac{f^{AB}_C j^C}{z_{12}},
    \\
    (j+b)^A(z_1) (j+b)^B(z_2) &\sim \frac{k\kappa^{AB}}{z_{12}^2} - \frac{k\kappa^{AB}}{z_{12}^2} + \frac{f^{AB}_C (j+b)^C}{z_{12}},
\end{align}
where $k=1$. In each choice of OPE channel to evaluate the $\mathbb{CP}^1$ current correlator in \eqref{eq: statement of single trace Parke-Taylor}, there can be only one appearance of $k\kappa^{A_1B_i}$, coming from choosing the first term in the $j^{A_1}(z_1) (j+b)^{B_i}(z_i)$ OPE. Every other appearance of $k \kappa^{AB}$ cancels pairwise. Although $k \neq 0$, $k$ only appears once (i.e as many times as the bare $j^A$ appears) in each OPE channel that contributes to the evaluation of a correlator. It is in this sense that $k^2=0$ for practical purposes, in the evaluation of the correlator at genus 0. 

\begin{figure} \label{fig: graph}
    \centering
    \includegraphics[width=0.4\linewidth]{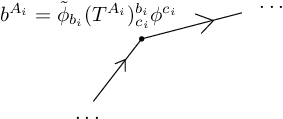}
    \caption{A single current insertion can be interpreted as a bivalent vertex in a directed graph. There is one incoming edge, where a $\tilde \phi$ from some other vertex can contract with the $\phi \in b^{A_i}$, and one outgoing edge, where the $\tilde \phi \in b^{A_i}$ can find a $\phi$ from some other vertex to contract with.}

    \vspace{1cm}
    
    \includegraphics[width=1\linewidth]{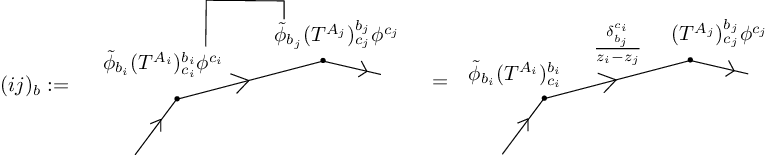}
    \caption{The Wick contraction of a $ \phi^{c_i}$ belonging to a current $b^{A_i}$ and the $\tilde \phi_{b_j}$ belonging to the current $b^{A_j}$ is the drawing of a single directed edge $(ij)_b$, where the $b$ subscript is a reminder that a boson pair was contracted.}
    \label{fig:directed edge}
\end{figure}

\paragraph{Notation for the proof and warmup} 
In order to introduce the diagrammatic language that will be useful later, it will be useful to introduce some notation. The free field realisation \cite{Wakimoto:1986gf} of $\mathfrak{g}_{1}$ , the Lie algebra $\mathfrak{g}$ at level $k=1$, is given by the bilinears $j^A$
\begin{align}
    &j^A(z) := \tilde \rho_b(T^A)^b_c \rho^c,
    \\
    &j^A(z_1) j^B(z_2) \sim \frac{\kappa^{AB}}{z_{12}^2} + \frac{f^{AB}_C j^C(z_2)}{z_{12}}
\end{align}
as can be verified using the free-field OPEs. Similarly, the free field realisation of $\mathfrak{g}_{-1}$, the Lie algebra $\mathfrak{g}$ at level $k=-1$, is given by the bilinears $b^A$ ($b$ for boson)
\begin{align}
    &b^A(z) :=  \tilde \phi_b(T^A)^b_c \phi^c,
    \\
    &b^A(z_1) b^B(z_2) \sim \frac{-\kappa^{AB}}{z_{12}^2} + \frac{f^{AB}_C b^C(z_2)}{z_{12}}
\end{align}
as can be verified using the free-field OPEs. As a warmup for the proof of equation \eqref{eq: statement of single trace Parke-Taylor}, we will introduce diagrammatics to demonstrate that correlators of $b^A$ or of $j^A$ evaluate to give a sum over products of trace factors
\begin{equation}
    \la \prod_{i=1}^n b^{A_i}(z_i) \ra =\sum_{\sigma \in S_n\setminus \mathbb{Z}_n} \frac{\text{tr}(T^{A_{\sigma_1}}\, \ldots \,T^{A_{\sigma_n}})}{z_{\sigma_1\sigma_2}\, \ldots \, z_{\sigma_n\sigma_1}} + \text{multi-trace contributions}
\end{equation}
This is easiest understood in the language of Feynman graphs and a diagrammatic enumeration of the many Wick contractions in the evaluation of the correlator. Let each bilinear $b^{A_i}$ correspond to a vertex $i$ in a graph with vertices $\{1,2, \ldots n\}$. Each $\tilde \phi \phi$ Wick contraction corresponds to a directed edge $(ij)_b$ (b for boson), where the right arrow means evaluation of the Wick contraction indicated (see Figure \ref{fig:directed edge})
\begin{align}
    &(ij)_b \rightarrow 
    \wick{
        {\tilde{\phi}_{b_i}}(z_i)(T^{A_i})_{c_i}^{b_i}\c1{\phi^{c_i}(z_i)}
        {\c1 {\tilde{\phi}_{b_j}}}(z_j)(T^{A_j})_{c_j}^{b_j}{\phi}^{c_j}(z_j)  
  }\nonumber
  \\
  &= \tilde \phi_{b_i}(z_i)(T^{A_i})_{c_i}^{b_i}
        \left(\frac{\delta^{c_i}_{b_j}}{z_i-z_j}\right)(T^{A_j})_{c_j}^{b_j} {{\phi}^{c_j}}(z_j)
\end{align}
Note that slings $(ii)_b$ are not permitted due to the assumed normal ordering of the bilinears $b^A$, and under e.g point splitting would evaluate to $0$ anyway due to tracelessness of the entries of the structure constants (recall that $G$ is semi-simple). Since each bilinear $b^{A_i}$ has one $\tilde \phi$ and one $\phi$, precisely one directed edge can end on the vertex $i$ and precisely one directed edge can start at the vertex $i$, for all $i$. Each vertex in the graph therefore has valence 2, with precisely one incoming and one outgoing edge. 
Each Feynman graph that contributes to the evaluation of a $\la \prod_i b^{A_i}(z_i)\ra$ current correlator is therefore the union of disjoint directed cycles. 

Compact notation for such a cycle (marked with the subscript $b$ for boson) is defined by
\begin{align}
    &(\sigma_1 \sigma_2 \ldots \sigma_m)_b:= (\sigma_1\sigma_2)_b(\sigma_2\sigma_3)_b...(\sigma_m\sigma_1)_b, \nonumber
    \\
    &\xrightarrow{\text{diagram evaluation}}  \frac{\text{tr}(T^{A_{\sigma_1}}\, \ldots \,T^{A_{\sigma_m}})}{z_{\sigma_1\sigma_2}\, \ldots \, z_{\sigma_m\sigma_1}}.
\end{align}
\begin{figure}
    \centering
    \includegraphics[width=0.7\linewidth]{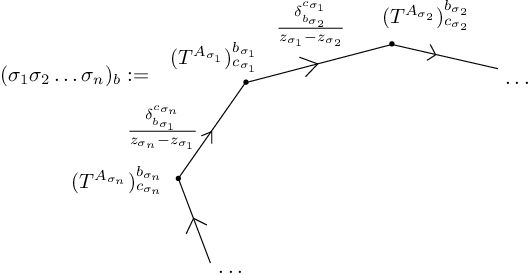}
    \caption{A directed cycle $(\sigma_1 \sigma_2 \ldots \sigma_m)_b:= (\sigma_1\sigma_2)_b(\sigma_2\sigma_3)_b...(\sigma_m\sigma_1)_b$ evaluates to a Parke-Taylor factor.}
    \label{fig:enter-label}
\end{figure}
The evaluation of the current correlator is therefore equivalent to summing over all ways of constructing sets of disjoint directed cycles from the $n$ vertices. 
\begin{equation}
    \la \prod_{i=1}^n b^{A_i}(z_i) \ra = \underbrace{\sum_{\sigma \in S_n\setminus \mathbb{Z}_n} (\sigma_1 \sigma_2 \ldots \sigma_n)_b}_{\text{single-trace}} + \frac{1}{2}\underbrace{\sum_{k=2}^{n-2}\left(\sum_{\sigma,\sigma'} (\sigma_1\ldots\sigma_k)_b(\sigma'_1\ldots\sigma'_{n-k})_b\right)}_{\text{double-trace}} + \ldots
\end{equation}
We have explicitly written the single trace term (where the sum is over non-cyclic permutations of the $n$ elements) and the double trace term, where the sums are over non-cyclic permutations of partitions of the $n$ elements into two sets. We define the compact cycle notation for particular Wick contractions of the fermionic bilinears $j^{A_i}$ in precisely the same way, with the subscript $f$ for fermion. The Feynman diagram evaluation of a fermion cycle comes with an overall $-$ sign compared to a boson cycle
\begin{align}
    &(\sigma_1 \sigma_2 \ldots \sigma_m)_f:= (\sigma_1\sigma_2)_f(\sigma_2\sigma_3)_f...(\sigma_m\sigma_1)_f  \nonumber
    \\ &\xrightarrow{\text{diagram evaluation}} -1 \times \frac{\text{tr}(T^{A_{\sigma_1}}\, \ldots \,T^{A_{\sigma_m}})}{z_{\sigma_1\sigma_2}\, \ldots \, z_{\sigma_m\sigma_1}}
\end{align}
This is due to the fermionic statistics of $\rho, \tilde \rho$ - the last $\tilde \rho, \rho$ contraction to be evaluated requires the $\tilde \rho$ and $\rho$ to exchange positions, incurring a minus sign from the anticommutation. A correlator of the fermionic bilinears $j^{A_i}$ is equivalent to summing over all ways of constructing sets of disjoint directed $f$ cycles from the $n$ vertices. 
\begin{equation}
    \la \prod_{i=1}^n j^{A_i}(z_i) \ra = \underbrace{\sum_{\sigma \in S_n\setminus \mathbb{Z}_n} (\sigma_1 \sigma_2 \ldots \sigma_n)_f}_{\text{single-trace}} + \frac{1}{2}\underbrace{\sum_{k=2}^{n-2}\left(\sum_{\sigma,\sigma'} (\sigma_1\ldots\sigma_k)_f(\sigma'_1\ldots\sigma'_{n-k})_f\right)}_{\text{double-trace}} + \ldots
\end{equation}
\paragraph{Proof of the single-trace assertion} Expanding the product in the first line of \eqref{eq: statement of single trace Parke-Taylor}, we see that the correlator is written as a sum of terms like
\begin{equation}
    \left\la j^{A_1}(z_1) \prod_{i=2}^n(b^{A_i}(z_i) + j^{A_i}(z_i)) \right\ra = \sum_{A \subset \{2,3,\ldots n\}}\left\la j^{A_1}(z_1) \prod_{i \in A}j^{A_i}(z_i)\right\ra\left\la \prod_{k \in \{2,3,\ldots n\}\setminus A}b^{A_k}(z_k)\right\ra
\end{equation}
The $j$ correlator gives a sum over all ways of forming sets of disjoint $f$ cycles from the elements of $\{1\} \cup A$, while the $b$ correlator gives a sum over all ways of forming disjoint $b$ cycles from the remaining unused elements, $\{2,3, \ldots n\}\setminus{A}$. The evaluation of the current correlator can be therefore written as a sum over a set of graphs $C$ 
\begin{equation}
    \left\la j^{A_1}(z_1) \prod_{i=2}^n(b^{A_i}(z_i) + j^{A_i}(z_i)) \right\ra = \sum_{c \in C} \underbrace{\text{Feyn}(c),}_{\text{the Feynman diagram evaluation of }c}
\end{equation}
\begin{figure}
    \centering
    \includegraphics[width=0.7\linewidth]{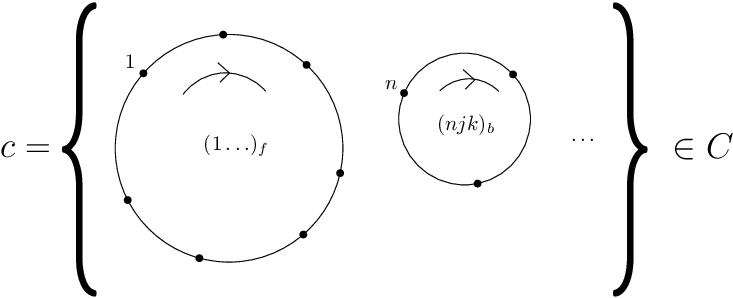}
    \caption{A cartoon depicting an example element $c \in C$. $C$ is the set of directed graphs with vertices $\{1, 2, \ldots ,n\}$ which are a union of disjoint directed $b, f$ cycles in which vertex 1 is required to be in a $f$ cycle. The curved arrow indicates the direction of the edges in the directed cycle, which is conventionally clockwise. The $\ldots$ indicate the other possible cycles that make up $c$ besides the two we have listed explicitly.}
    \label{fig:Disjoint cycles}
\end{figure}
where $C$ is the set of directed graphs with
\begin{enumerate}
    \item $n$ vertices labelled $1, \ldots, n$ each with exactly one incoming and one outgoing edge (i.e the union of disjoint directed cycles), 
    \item with each cycle labelled by a subscript $b$ or $f$, 
    \item with the constraint that the vertex $1$ is required to be in a cycle labelled subscript $f$.
\end{enumerate}
Each graph $c \in C$ has at least 1 $f$ cycle, since the vertex $1$ can only appear in a $f$ cycle, but it is perfectly allowed for a graph $c$ to have no $b$ cycles. An example element of $C$ is drawn in Figure \ref{fig:Disjoint cycles}. 

Define \textit{Zeilberger's involution} (inspired by \cite{ZEILBERGER198561}) $Z: C \rightarrow C$, in the following way.
\begin{enumerate}
    \item If $c$ is made up of a single cycle (this has to be a $f$ cycle, since vertex $1$ is required to be a participant of a $f$ cycle), do nothing
    \item If there is more than one cycle, consider the set of all bosonic or fermionic cycles that do not contain the $1$ vertex. Of these, pick the cycle with the highest numbered vertex and change its subscript (from $b$ to $f$ or vice versa).
\end{enumerate}
\begin{figure}
    \centering
    \includegraphics[width=0.7\linewidth]{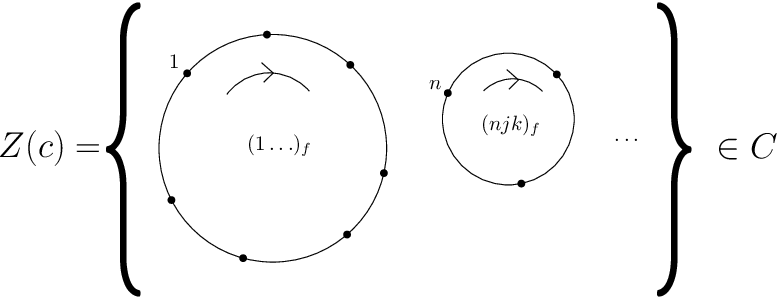}
    \caption{Taking Zeilberger's involution $Z$ on the cycle $c$ from Figure \ref{fig:Disjoint cycles} exchanges the subscript $b \leftrightarrow f$ of the cycle that has the highest numbered vertex and does not contain vertex $1$. In this case, it is the cycle $(njk)$, containing vertex $n$. If vertex $n$ were in the same cycle as vertex $1$, we would proceed in descending order (check which cycle vertex $n-1$ was in and try to exchange its suffix and so on) until we found the highest numbered vertex that was not in the same cycle as vertex $1$.}
    \label{fig:Z(Disjoint cycles)}
\end{figure}
Note that it immediately follows that $Z(Z(c))=c$. The action of Zeilberger's involution on the example element from Figure \ref{fig:Disjoint cycles} is given in Figure \ref{fig:Z(Disjoint cycles)}. The key point is that only the graphs $c = Z(c)$ in the stabilizer of Zeilberger's involution actually contribute to the evaluation of the current correlator, because we can show that the total contribution from all other terms vanishes. 

For all $c \neq Z(c)$, in the evaluation of the Feynman diagrams, the contribution of $\text{Feyn}(c)$ is equal and opposite to the contribution of $\text{Feyn}(Z(c))$. This is because $c, Z(c)$ only differ by the subscript label on one of their cycles. Due to the relative sign between bosonic and fermionic cycles, this means that $\text{Feyn}(c)+\text{Feyn}(Z(c))=0$ when $c \neq Z(c)$. 

Importantly, this means that the evaluation of the current correlator only receives contributions from the stabilizer of $Z$, which are precsiely the graphs that consist of a single fermion cycle. Therefore, we have the following
\begin{align}
    &\left\la j^{A_1}(z_1) \prod_{i=2}^n(b^{A_i}(z_i) + j^{A_i}(z_i)) \right\ra = \sum_{c \in C} \text{Feyn}(c) = \sum_{c \in \text{Stab}(Z)} \text{Feyn}(c) \nonumber
    \\
    &=-\sum_{\sigma \in S_n\setminus \mathbb{Z}_n} \frac{\text{tr}(T^{A_{\sigma_1}}\, \ldots \,T^{A_{\sigma_n}})}{z_{\sigma_1\sigma_2} \ldots z_{\sigma_n\sigma_1}}
\end{align}
and the assertion in equation \eqref{eq: statement of single trace Parke-Taylor} is proven. A simple corollary is
\begin{align}
     &\left\la b^{A_1}(z_1) \prod_{i=2}^n(b^{A_i}(z_i) + j^{A_i}(z_i)) \right\ra = \sum_{\sigma \in S_n\setminus \mathbb{Z}_n} \frac{\text{tr}(T^{A_{\sigma_1}}\, \ldots \,T^{A_{\sigma_n}})}{z_{\sigma_1\sigma_2} \ldots z_{\sigma_n\sigma_1}}
\end{align}

\section{Application to the Berkovits-Witten twistor string}\label{sec: application to the BW twistor string}
With the single-trace current system described above, genus 0 string correlators of the Berkovits-Witten (BW) twistor string \cite{Berkovits_2004} will compute pure super Yang-Mills processes, landing precisely on the RSVW formula \cite{Roiban_2004}. In this section, we go through in detail how this works and explain what it means for the tree-level effective field theory (EFT).

We begin with an introduction to the Berkovits-Witten twistor string. We replace the conventional current system with our single-trace current system, giving eq. \ref{eq: perturbative Yang-Mills sigma model} and discuss the evaluation of a genus 0 string correlator in subsection \ref{subsec: explicit evaluation of the worldsheet correlator}. Readers familiar with the Berkovits-Witten twistor string are encouraged to skip to section \ref{subsec: decoupling of csg}, where we discuss what the use of the single-trace current system means in terms of the string EFT.

For the purposes of this document, $\mathbb{PT}$ is the $\mathcal{N}=4$ \textit{supertwistor space}, the following open subset of $\mathbb{CP}^{3|4}$.
\begin{equation}
    \mathbb{PT} := \{Z^{\mathcal{M}}:=(Z^A, \psi^I)\in \mathbb{CP}^{3|4} | \lambda_{\alpha} = (\lambda_0,\lambda_1):=(Z^3,Z^4)\neq (0,0)\}
\end{equation}
The calligraphic $\mathcal{M} = A | I$ includes $A,I$ that each run from 1 to 4. $Z^{\mathcal{M}}$ is a homogenous coordinate on $\mathbb{CP}^{3|4}$, identified up to $\mathbb{C}^*$ rescalings. We label the latter two components of $Z^{A}$ as $\lambda_{\alpha}$. The locus removed from $\mathbb{CP}^{3|4}$ to get $\mathbb{PT}$ is the locus where $\lambda_{\alpha}$ vanishes.

The Berkovits-Witten twistor string as originally proposed \cite{Berkovits_2004} is an open string theory whose target space is twistor space, schematically written as 
\begin{equation}
    S_{BW}=\int_{\Sigma_{\text{open}}} Re(W_{\mathcal{M}}\bar D Z^{\mathcal{M}} + \text{ghosts} +  S_{\text{currents}})
\end{equation}
Vertex operators for gluons and conformal gravitons were constructed by Berkovits and Witten, and correlators of these vertex operators were computed at tree level. At tree level, the effective field theory (EFT) whose amplitudes are reproduced was found to be $\mathcal{N}=4$ super Yang-Mills coupled to some kind of $\mathcal{N}=4$ conformal supergravity, where the coupling to gravity was diagnosed precisely by the presence of the multi-color-traces in a genus 0 string correlator, i.e a tree-level scattering amplitude computation in the effective field theory.

Mason and Skinner \cite{Mason_2008} in 2008 and Reid-Edwards \cite{Reid-Edwards:2012vwx} in 2012 explained how to think about the Berkovits-Witten twistor string as an orientifold of a closed string theory (see the discussion starting at eq. 8.7 of \cite{Mason_2008}), and this is the perspective we will take because it streamlines notation and computations. In conformal gauge, the action takes the form of a chiral $\beta\gamma$ system
\begin{equation}\label{eq: Closed BW string}
    S_{\text{closed BW}}= \frac{1}{2}\left(\int_{\Sigma}W_{\mathcal{M}}\bar D Z^{\mathcal{M}} + \text{ghosts} + S_{\text{currents}}\right).
\end{equation}
We will defer the definition of the worldsheet fields to after this introductory preamble, when we define the model we want to work with in equation \eqref{eq: perturbative Yang-Mills sigma model}. $S_{\text{currents}}$ is a system of auxillary matter fields designed to cancel the conformal anomaly and to contain a family of spin-1 currents $j^A(z)$ that realise an affine Kac-Moody algebra for some gauge group $G$ at level $k$. As a quick reminder, in terms of local coordinates $z \in \mathbb{C}$ on a local patch of $\Sigma$, the defining OPE relations of $j^A(z)$ are
\begin{equation}
    j^{A_1}(z_1)j^{A_2}(z_2)\sim \frac{k \kappa^{A_1A_2}}{(z_1-z_2)^2}+\frac{f^{A_1A_2}_{A_3} j^{A_3}}{z_1-z_2}
\end{equation}
To go from the action in \eqref{eq: Closed BW string} back to the Berkovits-Witten open string, Mason and Skinner use a doubling trick. Define complex conjugates of all the worldsheet fields, completely constrained by the following orientifold condition
\begin{equation}
    \bar{\mathcal{O}}(z) = \mathcal{O}(\bar z)
\end{equation}
Dividing the closed worldsheet $\Sigma$ into two open string worldsheets $\Sigma_1,\Sigma_2 = \bar\Sigma_1$ exchanged by complex conjugation and glued along a common boundary, we find that the part of the Lagrangian $\mathcal{L}[\mathcal{O}]$ integrated over $\bar\Sigma_1$ can instead be rewritten as a $\bar{\mathcal{L}}[\bar{\mathcal{O}}]$ integrated over $\Sigma_1$
\begin{multline}
    S_{\text{closed BW}} = S_{\text{closed BW}}|_{\Sigma_1}+S_{\text{closed BW}}|_{\Sigma_2} = S_{\text{closed BW}}|_{\Sigma_1} + \bar S_{\text{closed BW}}|_{\Sigma_1}
    \\
    =\int_{\Sigma_1} Re(W_{\mathcal{M}}\bar D Z^{\mathcal{M}} + \text{ghosts} + S_{\text{currents}}),
\end{multline}
which is indeed the Berkovits-Witten open string action. In the closed worldsheet setting, introducing complex conjugate fields and fixing them using this complex conjugation can be interpreted as choosing a particular contour over which to integrate the complex-valued worldsheet fields. 
\begin{equation}
    \int \D \mathcal{O} \D\bar{\mathcal{O}}|_{\bar{\mathcal{O}}(z) = \mathcal{O}(\bar z)} = \int_{\Gamma} \D\mathcal{O},
\end{equation}
where $\Gamma$ is shorthand for the real cycle $\bar{\mathcal{O}}(z) = \mathcal{O}(\bar z)$ in the space of complex field configurations. This particular choice of contour that results from using the orientifold condition fixes the endpoints of the Berkovits-Witten open strings to be on $\mathbb{RP}^{3|4}\subset \mathbb{CP}^{3|4}$, but other choices of contours are possible.

The upshot is that by considering the closed string action \eqref{eq: Closed BW string} and imposing that the path integral over the complex valued fields is done over the contours $\Gamma$, a half-dimensional real cycle in the space of field configurations, we can work with a purely chiral theory that is equivalent to the Berkovits-Witten open string.

Constraining our discussion only to genus 0 and working in conformal gauge, the action takes the following form (where we have chosen to use the $\mathfrak{psl}(1|1)$ formalism for the single-trace current algebra)
\begin{align}\label{eq: perturbative Yang-Mills sigma model}
    S_{\text{pert. tree SYM}}&:=\int_{\mathbb{CP}^1} W_{\mathcal{M}}(\bar \partial +a)Z^{\mathcal{M}} + \int_{\mathbb{CP}^1} \tilde \phi_a \bar\partial \phi^a +  \tilde \rho_{a} \bar\partial \rho^{a} + u \bar \partial v\nonumber +(\text{aux. matter})_{c=26}
    \\
    &+\int_{\mathbb{CP}^1} \chi_+(\tilde \phi_a\rho^a + u) + \chi_-(\tilde \rho_a\phi^a+v(\tilde\phi_a \phi^a+\tilde \rho_a \rho^a)-N \partial v) + S_{\text{ghosts}}
\end{align}
To ensure that the total central charge vanishes, we require an additional $c=26$ to be supplied by a set of decoupled auxiliary matter fields. The worldsheet matter fields are
\begin{align}
    &Z^{\mathcal{M}} = (Z^M, \psi^I), \quad Z^M \in \Omega^{0}(\mathbb{CP}^1, \mathcal{L}_d), \quad \psi^I \in \Pi\Omega^{0}(\mathbb{CP}^1, \mathcal{L}_d), \quad M,I = 1, \ldots 4 \nonumber
    \\
    &W_{\mathcal{M}} = (W_M, \eta_I), \quad W_M \in \Omega^{0}(\mathbb{CP}^1, K_{\mathbb{CP}^1} \otimes (\mathcal{L}_d)^{-1}), \eta_I \in \Pi\Omega^{0}(\mathbb{CP}^1, K_{\mathbb{CP}^1} \otimes (\mathcal{L}_d)^{-1}),\nonumber
    \\
    &\tilde \rho_a \in \Pi\Omega^0(\mathbb{CP}^1, K^{1/2}_{\mathbb{CP}^1}), \, \rho^a \in \Pi\Omega^0(\mathbb{CP}^1, K^{1/2}_{\mathbb{CP}^1} ), \quad a = 1, \ldots, \underbrace{N}_{=\text{dim }F_{\mathfrak{g}}}\nonumber
     \\&\tilde \phi_a \in \Omega^0(\mathbb{CP}^1, K^{1/2}_{\mathbb{CP}^1}), \, \,\,\,\,\phi^a \in \Omega^0(\mathbb{CP}^1, K^{1/2}_{\mathbb{CP}^1}), \quad a = 1, \ldots, N \nonumber
     \\
     &     u \in \Pi\Omega^0(\mathbb{CP}^1, K_{\mathbb{CP}^1}), \, \,v \in \Pi\Omega^0(\mathbb{CP}^1)
\end{align}
and the gauge fields are $a \in \Omega^{0,1}(\mathbb{CP}^1), \chi_\pm \in \Pi\Omega^{0,1}(\mathbb{CP}^1)$. $Z^\mathcal{M}$ take values in a line bundle of degree $d \in \mathbb{Z}_{\geq 0}$ denoted $\mathcal{L}_d \cong \mathcal{O}(d)$, while $W_{\mathcal{M}}$ take values in the inverse. 

The current $W_\mathcal{M}Z^\mathcal{M}$ that generates $\mathbb{C}^*$ opposite rescalings of $W_\mathcal{M}, Z^\mathcal{M}$ is gauged. A convenient choice of gauge fixing for this symmetry is $a=0$, introducing a ghost system $m_2, n_2$
\begin{equation}
    S_{\text{twistor ghost}} = \int_{\mathbb{CP}^1} m_2\bar \partial n_2, \quad m_2 \in \Pi\Omega^0(\mathbb{CP}^1,K_{\mathbb{CP}^1}), \quad n_2 \in\Pi\Omega^0({\mathbb{CP}^1})
\end{equation}
We further gauge the $\mathfrak{psl}(1|1)$ currents just as in subsubsection \ref{subsubsec: psl gaugings}, such that the only bilinears of the free fields in the BRST cohomology are of the form $\tilde \phi_a\phi^b+\tilde \rho_a \rho^b$. The ghost systems associated to the gauge fixing $\chi_{\pm}=0$ are
\begin{align}
    S_{\text{ff ghost}} = \int_{\mathbb{CP}^1} \beta^+\bar \partial \gamma_+ + \beta^-\bar \partial \gamma_-,\quad \beta^\pm \in \Omega^0(\mathbb{CP}^1,K_{\mathbb{CP}^1}), \quad \gamma^\pm \in \Omega^0(\mathbb{CP}^1)
\end{align}
and the BRST operator is
\begin{equation}
    Q_{BRST}=\oint cT+n_2(W_{\mathcal{M}}Z^{\mathcal{M}}) + \gamma_+(\underbrace{\tilde \phi_a\rho^a}_{G^+} + u) + \gamma_-(\underbrace{\tilde \rho_a\phi^a}_{G^-}+v\underbrace{(\tilde\phi_a \phi^a+\tilde \rho_a \rho^a)}_{J}-N \partial v)
\end{equation}
The gauge-fixed action with the ghosts included therefore takes the form
\begin{align}
    S_{\text{pert. SYM gfixed}}&:=\int_{\mathbb{CP}^1} W_{\mathcal{M}}\bar \partial Z^{\mathcal{M}} + \int_{\mathbb{CP}^1} \tilde \phi_a \bar\partial \phi^a +  \tilde \rho_{a} \bar\partial \rho^{a} + \text{auxillary fields}\nonumber
    \\&+\int_{\mathbb{CP}^1} \underbrace{m_2 \bar \partial n_2 + \beta^+\bar \partial \gamma_+ + \beta^-\bar \partial \gamma_- + b \bar \partial c}_{\text{ghosts}}+ \underbrace{u \bar \partial v}_{\text{auxiliary free fermion}}
\end{align}
The ghosts and auxiliary free fermion in the last line decouple. By including one insertion of $v$ and one of $n_2$ in the evaluation of a correlator we saturate the fermionic zero modes and the ghost + auxiliary integrals can be done straightforwardly to give an overall multiple of 1.

The vertex operators for super Yang-Mills scattering states are twistor wavefunctions dressed with the currents $J^A$. For each external wavefunction $f_i(Z^\mathcal{M})\in H^{0,1}(\mathbb{PT})$, the fixed/integrated vertex operators are
\begin{align}\label{eq: fixed vertex operator}
    U^{A_i}(f_i)&:=c J^{A_i} f_i(Z^\mathcal{M}(z)) = c \delta(\gamma_+)\delta(\gamma_-) (\tilde \rho_a (T^{A_i})^a_b \rho^b) f_i(Z^\mathcal{M}(z))
    \\
    V^{A_i}(f_i)&:=\int_{\mathbb{CP}^1} J^{A_i} f_i(Z^\mathcal{M}(z)) = \int_{\mathbb{CP}^1} \delta(\gamma_+)\delta(\gamma_-) (\tilde \rho_a (T^{A_i})^a_b \rho^b) f_i(Z^\mathcal{M}(z))
\end{align}
As discussed in subsubsection \ref{subsubsec: psl gaugings}, the presence of the ghost delta functions $\delta(\gamma_+)\delta(\gamma_-)$ in $J^A$ means that we must insert $n-1$ picture changing operators $\Upsilon^\pm$ 
\begin{equation}
    \Upsilon^\pm := \delta(\beta^\pm)\delta_{Q_{BRST}}\beta^\pm
\end{equation}
and use them to descend $n-1$ of the $J^A$ currents. This is equivalent to computing with 1 $J^A$ and $n-1$ descended $\tilde J^A$ of the following form
\begin{equation}\label{eq: integrated vertex operator}
    \tilde J^{A_i}  =  (\tilde \phi_a (T^{A_i})^a_b \phi^b + \tilde \rho_a (T^{A_i})^a_b \rho^b)
\end{equation}
Precisely which vertex operator (of the 3 $U^A$s and the many $V^A$s) contains the lone $J^A$ current and which vertex operators contain the $\tilde J^A$ currents does not matter. The final step of the prescription is that a single insertion of an $v$ field is required at the location of the $J^A$ (recall the discussion after \eqref{eq: psl formalism correlator evaluaton}, that the $v$ must be placed at the location of a $\delta(\gamma_+)$).

The evaluation of genus 0 string correlators of these vertex operators is equivalent to the evaluation of a connected tree-level scattering amplitude in $\mathcal{N}  = 4$ super Yang-Mills.
\subsection{Explicit evaluation of the correlators}\label{subsec: explicit evaluation of the worldsheet correlator}
In this subsection we will remind ourselves how to explicitly evaluate the worldsheet correlators. We say `remind' because the actual evaluation is almost identical to the string correlator evaluation done by Berkovits \cite{Berkovits_2004}, except that our current algebra computes only a single-trace term, recovering the RSVW formula \cite{Witten_2004, Roiban_2004}
\begin{align}
    A_{\text{tree SYM}}&= \sum_{d \in \mathbb{Z}_{\geq 0}} g^{2d}\int \frac{\bigwedge_{k=0}^d \d^{4|4}Z_k }{\text{Vol } GL(2, \mathbb{C})}\frac{\bigwedge_{i=1}^n   f_i(Z(\sigma_i))(\sigma_i \d \sigma_i)}{(\sigma_{p_0(1)}\sigma_{p_0(2)})\ldots (\sigma_{p_0(n)}\sigma_{p_0(1)})}
\end{align}
precisely from the evaluation of the gauge-fixed, descended, and ghost-integral-performed correlator $\mathcal{C}_{\text{tree}}$ 
\begin{align}
     \mathcal{C}_{\text{tree}}:=&\frac{1}{\text{Vol } GL(2, \mathbb{C})} \sum_{d \in \mathbb{Z}_{\geq 0}} g^{2d} \frac{1}{Z} \int \D (\text{fields}) \exp\left( \int_{\mathbb{CP}^1_\sigma}W_{\mathcal{M}} \bar \partial Z^{\mathcal{M}} +  \tilde \phi \bar \partial \phi + \tilde \rho \bar \partial \rho \right) \times  \nonumber
    \\
    &\int_{\mathbb{CP}^1_{\sigma}} f_1(Z(\sigma)) j^{A_1}(\sigma) \prod_{i=2}^n  \int_{\mathbb{CP}^1_{\sigma}} f_i(Z(\sigma)) (b^{A_i}(\sigma)+j^{A_i}(\sigma))
    \nonumber
\end{align}
In this section we revert to using homogenous coordinates $\sigma^a$ on the worldsheet $\mathbb{CP}^1$. To distinguish worldsheet coordinates from twistor or kinematic spinors $\lambda^\alpha$, we use round brackets to indicate contraction $(\sigma_1\sigma_2):=\sigma_1^a \sigma_{2a}$. 

First we see that the evaluation of the path integral factorises into a piece that depends on $\tilde \phi, \phi, \tilde \rho, \rho$, and a piece that depends on the twistor variables. Using the result of subsubsection \ref{subsec: single trace correlator evaluation}, we can directly evaluate the $\tilde \phi, \phi, \tilde \rho, \rho$ path integral.
\begin{equation}
    \la j^{A_1}(\sigma_1) \prod_{i=2}^n (b+j)^{A_i}\ra = -\sum_{p \in S_n}\frac{\text{tr}(T^{A_{p(1)}}\ldots T^{A_{p(n)}})}{(\sigma_{p(1)}\sigma_{p(2)})\ldots (\sigma_{p(n)}\sigma_{p(1)})}\bigwedge_{i=1}^n (\sigma_i \d \sigma_i) 
\end{equation}
As is usual, we will write $\mathcal{C}_{\text{tree}}$ as a sum over color orderings
\begin{equation}
    \mathcal{C}_{\text{tree}} =: \sum_{p \in S_n} \text{tr}(T^{A_{p(1)}}\ldots T^{A_{p(n)}}) \mathcal{C}_{\text{tree}, p}
\end{equation}
From this definition, the expression for $C_{\text{tree}, p_0}$ for some $p_0 \in S_n$ is therefore
\begin{align}
    \mathcal{C}_{\text{tree }p_0}&=\frac{-1}{\text{Vol } GL(2, \mathbb{C})} \sum_{d \in \mathbb{Z}_{\geq 0}} g^{2d} \frac{1}{Z} \int \D (\text{fields}) \exp\left( \int_{\mathbb{CP}^1_\sigma}W_{\mathcal{M}} \bar \partial  Z^{\mathcal{M}}\right) \times  \nonumber
    \\
    &\frac{\bigwedge_{i=1}^n  \int_{\mathbb{CP}^1_{\sigma}} f_i(Z(\sigma_i))(\sigma_i \d \sigma_i)}{(\sigma_{p_0(1)}\sigma_{p_0(2)})\ldots (\sigma_{p_0(n)}\sigma_{p_0(1)})} 
    \nonumber
\end{align}
As is conventional, we have written the correlator in terms of only the integrated vertex operators, capturing the effect of the three $c$ ghost insertions by writing
\begin{equation}
    \frac{1}{\text{Vol} SL(2, \mathbb{C})}:= \frac{(\sigma_1\sigma_2)(\sigma_2 \sigma_3)(\sigma_3 \sigma_1)}{(\sigma_1 \d \sigma_1)(\sigma_2 \d \sigma_2)(\sigma_3 \d \sigma_3)}
\end{equation}
and an additional dividing out by $ \text{Vol} \mathbb{C}^*$ to account for the reundancy of constant overall rescalings of $Z^\mathcal{A}$. Integrating out $W_{\mathcal{M}}$ to learn holomorphy of $Z^\mathcal{M}$, we have
\begin{align}
    \mathcal{C}_{\text{tree }p_0}&=\frac{-1}{\text{Vol } GL(2, \mathbb{C})} \sum_{d \in \mathbb{Z}_{\geq 0}} g^{2d}\int \bigwedge_{k=0}^d \d^{4|4}Z_k \frac{\bigwedge_{i=1}^n  \int_{\mathbb{CP}^1_{\sigma}} f_i(Z(\sigma_i))(\sigma_i \d \sigma_i)}{(\sigma_{p_0(1)}\sigma_{p_0(2)})\ldots (\sigma_{p_0(n)}\sigma_{p_0(1)})} 
    \nonumber
\end{align}
This gives the RSVW formula as desired.
\subsection{Decoupling of $\mathcal{N}  = 4$ conformal supergravity intermediates at tree-level} \label{subsec: decoupling of csg}
In the Berkovits-Witten twistor string, the presence of the multi-trace terms in the connected tree-level gluon amplitude is understood in terms of Feynman diagrams in which $\mathcal{N}  = 4$ conformal supergravity intermediate states propagate between pure $\mathcal{N}=4$ super Yang-Mills Feynman diagrams. How does the single-trace current system decouple these intermediates? Although we have already seen at a nuts-and-bolts level how this is accomplished from a worldsheet correlator evaluation, it is enlightening to see the interpretation in terms of the effective field theory.

We will only need the $k$-dependence of the EFT action for the following argument, so the precise definitions are deferred to Appendix \ref{appendix: BW twistor string}. Tree-level correlators computed by the Berkovits-Witten twistor string (where the currents used are a conventional Kac-Moody at level $k$ rather than the single-trace current system) are consistent with tree amplitudes computed with the following twistor action (the precise form of the interaction term was first determined by Mason and Skinner \cite{Mason_2008}, see eq. 73 and the unnumbered following eq.),
\begin{multline}\label{eq: EFT for BW}
    S_{BW}=\int \D^{3|4}Z \wedge \left(k \times hCS(\mathcal{A})-hCS(\partial_{\mathcal{M}} f^{\mathcal{N}}) + N(f) \lrcorner g\right) \\ + \int \d^{4|8}x \exp\left[\left(\int_{\mathcal{L}_x} g \right)+ k\log \det(\bar \partial + \mathcal{A}) - \log \det(\bar \partial + {\partial_{\mathcal{M}} f^{\mathcal{N}}})\right],
\end{multline}
$\mathcal{A}$ is the $\mathcal{N}=4$ super Yang-Mills superfield, while $f,g$ are superfields that encode the $\mathcal{N}=4$ conformal supergravity degrees of freedom. $hCS(\mathcal{A})$ is the holomorphic-Chern-Simons 3-form,
\begin{equation}
    hCS(\mathcal{A}):=\int \D^{3|4}Z \wedge \text{tr}\left(\mathcal{A} \bar D \mathcal{A} + \frac{2}{3}\mathcal{A}^3\right), \quad \quad \bar D:= \bar \partial + \mathcal{L}_{f \cdot \partial}
\end{equation}
$\int_{\mathcal{L}_x}$ means an integral over a curved twistor line, and the $\log \det$ can be written explicitly in terms of frame fields.
\begin{equation}
    \log \det(\bar \partial + \mathcal{A}) := \int_{\mathcal{L}_x \times \mathcal{L}_x} \frac{\D\sigma_1\D\sigma_2}{( \sigma_1 \sigma_2)^2} \text{tr}\left(H\mathcal{A}H^{-1}(Z(\sigma_1))H\mathcal{A}H^{-1}(Z(\sigma_2))\right), \quad \bar \partial_{\mathcal{L}_x} H = \mathcal{A}|_{\mathcal{L}_x}H. 
\end{equation}
The interested reader is directed to Appendix \ref{appendix: BW twistor string} for the full details.

For the discussion that follows, the only salient feature of the Berkovits-Witten EFT is the prefactor of $k$ (the level of the worldsheet current algebra) in both the $hCS(\mathcal{A})$ and in the $\log \det$ interaction term. Its appearance in the EFT can be deduced from the string correlators by counting traces - each color trace in a Berkovits-Witten string correlator\footnote{following a very similar argument to that outlined in and around eq. \ref{eq: traces weighted with k}} comes with one power of $k$. Going through the Feynman rules of the EFT, we find that the EFT must have had each color trace weighted with a factor of $k$.

How does the modified action \eqref{eq: perturbative Yang-Mills sigma model} remove the contributions of conformal graviton exchange to the evaluation of a tree level gluon scattering amplitude? The key fact is that correlator computations with the action \eqref{eq: perturbative Yang-Mills sigma model} give a subsector of the following \textit{different} theory
\begin{align}\label{eq: single trace system EFT}
    S[\mathcal{A}_1,\mathcal{A}_2,f,g] &= \int \D^{3|4}Z \wedge \left[\sum_{i=1}^2(k_i \times hCS(\mathcal{A}_i))-hCS(\partial_I f^J) + N(f) \lrcorner g\right] 
    \\ &+ \int \d^{4|8}x \exp\left[\int_{\mathcal{L}_x} g + (\sum_{i=1}^2(k_i \log \det(\bar \partial + \mathcal{A}_i)) - \log \det(\bar \partial + {\partial_I f^J})\right],
    \\
    &\mathcal{A}_1,\mathcal{A}_2 \in \Omega^{0,1}(\mathbb{PT}, \mathfrak{g}), \quad k_1=1, \quad k_2=-1
    \\
    &f \in \Omega^{0,1}(\mathbb{PT}, T^{1,0}\mathbb{PT}), \quad g \in \Omega^{1,1}(\mathbb{PT})
\end{align}
One perspective on the action \eqref{eq: single trace system EFT} is that we start with a single copy of super Yang-Mills with gauge group $GL(N|N,\mathbb{C})$ and set to zero every gluon that was not valued in $\mathfrak{g} \oplus \mathfrak{g} \subset \mathfrak{gl}(N,\mathbb{C}) \oplus \mathfrak{gl}(N,\mathbb{C}) \subset \mathfrak{gl}(N|N,\mathbb{C})$
\begin{equation}
    \mathcal{A}_{\mathfrak{gl}(N|N)} = \begin{pmatrix}
        \mathcal{A}_2 & 0_N \\ 0_N & \mathcal{A}_1
    \end{pmatrix}, \quad hCS(\mathcal{A}_{\mathfrak{gl}(N|N)}) = hCS(\mathcal{A}_1)-hCS(\mathcal{A}_2),
\end{equation}
where the relative sign between the two terms comes from the relative sign in the supertrace. In any case, it can be read off that there are 2 copies of $\mathcal{N}  = 4$ super Yang-Mills (coupled to 1 copy of conformal supergravity), where one copy has $k=1$ and the other copy has $k=-1$. 

The reason this is the case is that in the worldsheet theory \eqref{eq: perturbative Yang-Mills sigma model}, in the absence of the $\mathfrak{psl}(1|1)$ gauge fixing, currents from the free boson bilinears and the free fermion bilinears would allow us to define two distinct types of super Yang-Mills vertex operator, one with level $k=1$ (corresponding to the field theory calculations with $\mathcal{A}_1$) and the other with level $k=-1$ (corresponding to the field theory calculations with $\mathcal{A}_2$).
\begin{align}
    \mathcal{A}_1 \text{ scattering state} = \int \tilde \rho T \rho f_i(Z)
    \\
    \mathcal{A}_2  \text{ scattering state} =\int \tilde \phi T \phi f_i(Z)
\end{align}

However, we do have the $\mathfrak{psl}(1|1)$ gauge fixing, meaning that the worldsheet vertex operators we are allowed to use in each nonzero correlator are constrained. In terms of the field theory, this means that we are in a subsector of the theory in which the allowed field theory correlators must include one $\mathcal{A}_1$, any number of $\mathcal{A}_1+\mathcal{A}_2$, but no lone $\mathcal{A}_2$ insertions. The decoupling of the conformal supergravity intermediates from the computation of a connected all-external-gluon tree-level process follows from an argument identical in spirit to the Zeilberger involution argument of subsubsection \ref{subsec: single trace correlator evaluation}.

\paragraph{Lemma} Connected tree level amplitudes computed by LSZ reduction of the correlators $\la \mathcal{A}_1(Z_1) \prod_{i=2}^n (\mathcal{A}_1+\mathcal{A}_2)(Z_i) \ra$ computed with the EFT action \eqref{eq: single trace system EFT} give connected pure $\mathcal{N}  = 4$ super Yang-Mills tree amplitudes.
\paragraph{Proof} First, notice that the wrong-sign kinetic term in $hCS(A_2)$ and the wrong-sign interaction term for $\mathcal{A}_2$ means that connected tree-level pure super Yang-Mills Feynman diagrams computed using the Feynman rules for $\mathcal{A}_2$ all come with a relative minus sign compared to those of $\mathcal{A}_1$. 

This is because the number of edges of a connected tree is precisely one less than the number of vertices. Therefore, in the construction of a connected tree Feynman diagram for a pure super Yang-Mills process of $\mathcal{A}_2$, each additional use of the interaction vertex comes with one additional use of the propagator. This means that every connected tree diagram constructed for $\mathcal{A}_2$ has a consistent relative sign compared to the equivalent connected tree diagram constructed for $\mathcal{A}_1$. The relative sign is $-1$, from the base case where the tree diagram is a single interaction vertex.

Every tree-level gluon amplitude computed using the Feynman rules of the action \eqref{eq: single trace system EFT} comes from gluing together pure super Yang-Mills processes from either $\mathcal{A}_1$ or $\mathcal{A}_2$ with conformal supergravity propagators. Since $f,g$ couple to the interaction terms of both $\mathcal{A}_1$ and of $\mathcal{A}_2$, this means that conformal supergravity propagators can run between a pure super Yang-Mills process of $\mathcal{A}_i$ to a pure super Yang-Mills process of $\mathcal{A}_j$, for $i,j = 1,2$.

In the computation of a correlator of $\la \mathcal{A}_1(Z_1) \prod_{i=2}^n (\mathcal{A}_1+\mathcal{A}_2)(Z_i) \ra$, there is at least one pure super Yang-Mills tree of $\mathcal{A}_1$ in each Feynman diagram. Define Zeilberger's involution on Feynman diagrams, in which the super Yang-Mills process that $\mathcal{A}(Z_r)$ participates in is exchanged from $\mathcal{A}_1$ to $\mathcal{A}_2$ or vice versa, where $r$ is the highest number such that $\mathcal{A}(Z_r)$ does not participate in the same pure super Yang-Mills process as does $\mathcal{A}(Z_1)$.

Because of the relative $-$ sign between a connected tree-level pure super Yang-Mills Feynman diagram of $\mathcal{A}_1$ and the equivalent Feynman diagram of $\mathcal{A}_2$, we find that only the stabilizer of Zeilberger's involution contributes to the tree-level evaluation of the correlator. These are the Feynman diagrams where there is a single pure Yang-Mills process of $\mathcal{A}_1$ $\square$.

\paragraph{Limitations} It is explicit that this proof relied on there being a consistent relative $-$ sign between the connected tree-level pure super Yang-Mills processes of $\mathcal{A}_1$ and the equivalent for $\mathcal{A}_2$. When the number of edges plus the number of vertices of the Feynman graph is no longer odd, such as for a connected 1 loop pure Yang-Mills process, the diagrams for $\mathcal{A}_1$ and $\mathcal{A}_2$ have the same sign, and add. In this case, there is no mechanism with which to exclude diagrams that have conformal supergravity intermediates. Beyond tree-level, at the level of the string EFT it is clear that this proposed mechanism for decoupling conformal supergravity and having only pure super Yang-Mills does not work. We hope that by moving away from a string interpretation, a twistor worldsheet theory for pure $\mathcal{N}  = 4$ super Yang-Mills can still be constructed, as we expand on in the discussion \ref{sec: Discussion}.

\subsection{Collinear limits and splitting function}\label{subsec: collinear limits and splitting function}
A major advantage of using the single-trace current system in the Berkovits-Witten twistor string (instead of a conventional affine Kac-Moody at level $k$) is that we can perform a direct computation of the collinear limits and splitting functions of $\mathcal{N}  = 4$ super Yang-Mills tree amplitudes using the OPE rules of the worldsheet correlator evaluation. Trying to use a conventional affine Kac-Moody would result in unphysical double poles (interpreted in this setting as conformal graviton exchange). As expected, the results agree with those obtained by using a conventional affine Kac-Moody with the level $k$ set to zero by hand \cite{Adamo:2021zpw, Bu:2021avc}.

To set the scene, recall that the twistor wavefunctions for gluon scattering take the form \cite{Witten_2004, Berkovits_2004}
\begin{equation}
    (C )_Af (Z(\sigma))=(C )_A\int \frac{\d t}{t} \bar \delta^2(t \lambda^\alpha (\sigma) - \sqrt\omega \kappa ^\alpha)e^{t\sqrt\omega ([\mu(\sigma) \tilde \kappa ]+\psi(\sigma) \cdot \eta )}
\end{equation}
The external data in this plane wave are $(C )_A, \omega , (\kappa ^\alpha, (\tilde \kappa )_{\dot \alpha}, (\eta )_I)$. $C_A$ is a constant vector in color space. To understand what $\omega , (\kappa ^\alpha, (\tilde \kappa )_{\dot \alpha}, (\eta )_I)$ correspond to on spacetime, we can take the pullback of $f (Z)$ to a twistor line $L_{(x,\theta)}=\{\mu^{\dot \alpha} = ix^{\alpha \dot \alpha}\lambda_\alpha, \quad \psi^I = \theta^{\alpha I}\lambda_\alpha\}$
\begin{align}
    &f |_{L_{(x,\theta)}} = \int \frac{\d t}{t} \bar \delta^2(t \lambda^\alpha (\sigma) - \sqrt\omega \kappa ^\alpha)e^{ix \cdot P +\theta \cdot \Pi }
    \\
    &P ^{\alpha \dot \alpha} := \omega  \kappa ^\alpha \tilde \kappa ^{\dot \alpha}, \quad (\Pi )^{\alpha}_I := \omega \kappa ^\alpha \eta_I
\end{align}
We can read off that $P , \Pi $ are the external momenta that characterise a scattering state (i.e eigenvalues under translations of $x, \theta$ respectively). Since $V(f ) = \int f (b+j)$, the $\mathbb{CP}^1$ integral in the definition of the vertex operator can be done against the delta function in the definition of the external scattering wavefunction $f $. To simplify notation, we introduce the following notation for the components of the sections of powers of $K_{\mathbb{CP}^1}$
\begin{equation}
    o(\sigma) =: O(\sigma) \sqrt{(\sigma \d \sigma)}, \quad \begin{pmatrix}
        o(\sigma) \\ O(\sigma)
    \end{pmatrix} \in \begin{pmatrix}
        K^{1/2}_{\mathbb{CP}^1} \\ \mathcal{O}(-1)
    \end{pmatrix}
\end{equation}
where $o \in \{\tilde \phi, \phi, \tilde \rho, \rho\}$ and $O \in \{\tilde \Phi, \Phi, \tilde P, P\}$\footnote{Capital $\rho$ is $P$ in the Greek alphabet.}. In terms of the component fields, the free field Wick contraction rules of $\tilde \phi, \phi$ and $\tilde \rho, \rho$ are written as
\begin{equation}
    \wick{ \c1{\tilde \Phi_a(\sigma_1)} \c1{\Phi^b(\sigma_2)}} = \frac{-\delta_a^b}{(\sigma_1 \sigma_2)}, \quad \wick{ \c1{\tilde P_a(\sigma_1)} \c1{P^b(\sigma_2)}} = \frac{\delta_a^b}{(\sigma_1 \sigma_2)}
\end{equation}
It will also be convenient to name the components of the $b^A, j^A$ bilinears, which we will call $B^A, J^A$
\begin{align}
    b^A(\sigma) &=: B^A(\sigma) (\sigma \d \sigma), \quad B^A(\sigma) =  \tilde \Phi_b(T^A)^b_c \Phi^c  \in \Omega^0(\mathbb{CP}^1_\sigma, \mathcal{O}(-2))
    \\
    j^A(\sigma) &=: J^A(\sigma) (\sigma \d \sigma), \quad J^A(\sigma)  = \tilde P_b (T^A)^b_c P^c\in \Omega^0(\mathbb{CP}^1_\sigma, \mathcal{O}(-2))
\end{align}
Written in terms of these components, the Wick contractions between the $B^A$ gives
\begin{align}\label{eq: Wick contractions for B}
    B^{A_1}(\sigma_1)B^{A_2}(\sigma_2) &= \frac{-\kappa^{A_1 A_2}}{(\sigma_1\sigma_2)^2} + \frac{\Phi^b(\sigma_1)(T^{A_1})^c_b(T^{A_2})^d_c\tilde \Phi_d(\sigma_2) -\Phi^b(\sigma_2)(T^{A_2})^c_b(T^{A_1})^d_c\tilde \Phi_d(\sigma_1)}{(\sigma_1 \sigma_2)} \nonumber
    \\ &+ :B^{A_1}(\sigma_1)B^{A_2}(\sigma_2):
\end{align}
This is a strict equality rather than an OPE limit, where we compile the terms in which Wick contractions are taken between $B^{A_1}$ and $B^{A_2}$. In the OPE limit $(\sigma_1 \sigma_2) \rightarrow 0$, we have $\sigma^a_1 \rightarrow \sigma^a_2 \frac{(\sigma_1 \nu)}{(\sigma_2 \nu)}$ for a constant reference spinor $\nu^a$, and we may make the replacements $O(\sigma_1) \rightarrow O(\sigma_2)\frac{(\sigma_2 \nu)}{(\sigma_1 \nu)}$ for sections of $\mathcal{O}(-1)$. The OPE limit therefore gives
\begin{equation}
    B^{A_1}(\sigma_1)B^{A_2}(\sigma_2) \sim \frac{-\kappa^{A_1 A_2}}{(\sigma_1\sigma_2)^2} + \frac{f^{A_1A_2}_{A_3} B^{A_3}(\sigma_2)}{(\sigma_1 \sigma_2)} \frac{(\sigma_2 \nu)}{(\sigma_1 \nu)}
\end{equation}
The same is true for $J^A$,
\begin{align}
    J^{A_1}(\sigma_1)J^{A_2}(\sigma_2) \sim \frac{\kappa^{A_1 A_2}}{(\sigma_1\sigma_2)^2} + \frac{f^{A_1A_2}_{A_3} J^{A_3}(\sigma_2)}{(\sigma_1 \sigma_2)} \frac{(\sigma_2 \nu)}{(\sigma_1 \nu)}
\end{align}
The vertex operator for the $i$th external gluon scattering state (in the case of a fully descended vertex operator) with a constant color vector contracted is written in this notation as
\begin{align}
    C_i \cdot  V(f_i) &=\int_{\mathbb{CP}^1} f_i(Z(\sigma))(C_i)\cdot(b+j)(\sigma) \nonumber
    \\
    &=\int_{\mathbb{CP}^1}(\sigma \d \sigma) (C_i)_A(B^A(\sigma) + J^A(\sigma))\int \frac{\d t}{t} \bar \delta^2(t \lambda^\alpha (\sigma) - \sqrt\omega_i\kappa_i^\alpha)e^{t\sqrt\omega_i([\mu(\sigma) \tilde \kappa_i]+\psi(\sigma) \cdot \eta_i)}
\end{align}
In order to do this evaluation, it is most convenient to merge the $\d t$ integral over $\mathbb{C}^*$ and the integral over $\mathbb{CP}^1$ into an integral over $\mathbb{C}^2 \setminus (0,0)$ coordinatised by $w^a := \sigma^a t^{1/d} \omega_i^{-1/2d}$. The holomorphic measures are related by
\begin{equation}
    (\sigma \d \sigma) \frac{\d t}{t} (B(\sigma) + J(\sigma))^A= \frac{d}{2} \d^2 w (B(w) + J(w))^A
\end{equation}
This reparameterisation of the variables of integration allows us to conveniently evaluate the delta function
\begin{align}\label{eq: vertex operators evaluated}
    C_i \cdot  V(f_i) &= \frac{d}{2\omega_i}\int_{\mathbb{C}^2\setminus (0,0)} \d ^2 w \,\, \bar \delta^2(\lambda^\alpha (w) - \kappa_i^\alpha)e^{\omega_i([\mu(w) \tilde \kappa_i]+\psi(w) \cdot \eta_i)} C_i\cdot (B + J)(w)\nonumber
    \\
   & =\frac{d}{2\omega_i}\sum_{k=1}^d\left(\frac{e^{\omega_i([\mu(w) \tilde \kappa_i]+\psi(w) \cdot \eta_i)} C_i\cdot (B+J)(w)}{\det \left(\frac{\partial \lambda^\alpha}{\partial w^b}\right)} \right)_{w = I_k},\quad \lambda^\alpha(I_k) -\kappa_i^\alpha = 0 \,\,\forall k
\end{align}
The result is a sum over the $d$ solutions labelled $w^a = I^a_k, k \in \{1, \ldots d\}$, which for generic degree $d$ maps $\lambda^\alpha(\sigma)$ is nondegenerate. The Jacobian factor in the denominator comes from solving against the delta functions.

\subsubsection{Holomorphic Collinear limits}
Taking the holomorphic collinear limit of particles $i, j$ in a connected $\mathcal{N}  = 4$ super Yang-Mills scattering amplitude means to consider the singularity structure of the amplitude in the limit $\la \kappa_i \kappa_j \ra \rightarrow 0$, where external momenta are complexified. We are interested in the leading singularities, which start with a simple pole for the tree level amplitude.
\begin{equation}
    A^{\text{$\mathcal{N}$  = 4 SYM}}_{\text{conn. tree}}(f_1, \ldots f_n) = \frac{1}{\la \kappa_i \kappa_j \ra}(\ldots) + \mathcal{O}({\la \kappa_i \kappa_j \ra^0})
\end{equation}
We can investigate the singularities in the connected super Yang-Mills tree amplitude from the point of view of the worldsheet model by asking what happens when we consider the singularity structure implied by the correlator evaluation when we take $\la \kappa_i \kappa_j \ra \rightarrow 0$
\begin{equation}
    \la \ldots C_i \cdot  V(f_i) \ldots C_j \cdot  V(f_j)\ldots \ra_{\text{pert. SYM}} = \frac{1}{\la \kappa_i \kappa_j \ra}(\ldots) + \mathcal{O}({\la \kappa_i \kappa_j \ra^0})
\end{equation}
In the limit $\la \kappa_i \kappa_j \ra \rightarrow 0$, generically each preimage of $\kappa_i$ under the map $\lambda$ will approach one preimage of $\kappa_j$. Without loss of generality, assign subscripts $k \in \{1, \ldots d\}$ such that the preimage $I_k\in \lambda^{-1}(\kappa_i)$ approaches preimage $J_k\in \lambda^{-1}(\kappa_j)$ for each $k$. Then the terms in $V(f_i)$ given as the sum over insertions at each $[I_k] \in \mathbb{CP}^1$ will give singularities with the terms in $V(f_j)$ when they Wick contract. 
\begin{align}
    C_i \cdot  V(f_i) C_j \cdot V(f_j) &\sim \frac{d^2}{4\omega_{i}\omega_j}\sum_{k=1}^d\left(\frac{e^{([\mu(w)  (\omega_i\tilde\kappa_i+\omega_j\tilde\kappa_j)]+\psi(w) \cdot (\omega_i\eta_i+\omega_j\eta_j))} (f C_iC_j)\cdot (B+J)(w)}{(I_kJ_k)\left(\det \left(\frac{\partial \lambda^\alpha}{\partial w^\beta}\right)\right)^2} \right)_{w = I_k} \nonumber
    \\
    &+ \text{regular}
\end{align}
We have given the result for two vertex operators each with a fully descended $b +j$ current. If one had a $\delta(\gamma_+)\delta(\gamma_-)j$ current, the result is identical except $B+J$ must be replaced by $\delta(\gamma_+)\delta(\gamma_-)J$. 

The fact that the double pole from the double contractions between $V(f_i),V(f_j)$ for $i,j \in \{1,\ldots,n\}$ does not contribute to the evaluation of the correlator\footnote{For $i,j \in \{2,\ldots,n\}$ this follows from a direct computation, since the contribution from the bosonic double contraction cancels that of the fermionic double contraction.} is a corollary of the result from \ref{subsec: single trace correlator evaluation}. Such a double contraction would result in a disconnected contribution, i.e a bubble diagram (2 particle trace) multiplying the rest of the correlator evaluation. But we know that these processes do not contribute to the correlator evaluation, as they are not in the stabilizer of Zeilberger's involution.

To relate the $1/(I_k J_k)$ pole from Wick contraction to a putative $1/\la \kappa_i \kappa_j\ra$ pole, we have the useful identity
\begin{align}
    &\lim_{\la \kappa_i \kappa_j\ra\rightarrow 0}\frac{\la \kappa_i \kappa_j\ra}{(I_k J_k)} = \lim_{\la \kappa_i \kappa_j\ra\rightarrow 0}\frac{\la \lambda(I_k) \lambda(J_k)\ra}{(I_k J_k)} \nonumber
    \\
    &=\lim_{\la \kappa_i \kappa_j\ra\rightarrow 0}\frac{1}{(I_k J_k)}\left(\underbrace{\left(\frac{1}{d} I_k^b\frac{\partial \lambda^\alpha}{\partial I_k^b}\right)}_{\lambda^\alpha(I_k)} \underbrace{\left(\lambda_{\alpha}(I_k) +\left(J_k-I_k, \frac{\partial}{\partial I_k}\right)\lambda_\alpha(I_k) +\ldots\right)}_{\lambda_\alpha(J_k)} \right)\nonumber
    \\
    &= \frac{-1}{2d} \det \left(\frac{\partial \lambda^\alpha}{\partial w^b}\right)_{w=I_k=J_k}
\end{align}
In going to the 3rd line we have used the fact that the two copies of $\partial \lambda/\partial I_k$ are skew on one index, and therefore can be written as contracted on both indices.
\begin{equation}
    \frac{\partial \lambda^\alpha}{\partial I_k^b} \frac{\partial \lambda_\alpha}{\partial I_k^c} = \frac{-\epsilon_{bc}}{2} \det \left(\frac{\partial \lambda}{\partial I_k}\right)
\end{equation}
We therefore find that the singularities between $V(f_i) V(f_j)$ as $\la \kappa_i \kappa_j \ra \rightarrow 0$ are
\begin{align}
    C_i \cdot \tilde V(f_i) C_j \cdot\tilde V(f_j) &\sim \frac{-d}{8\omega_{i}\omega_j \la \kappa_i \kappa_j\ra}\sum_{k=1}^d\left(\frac{e^{([\mu(w)  (\omega_i\tilde\kappa_i+\omega_j\tilde\kappa_j)]+\psi(w) \cdot (\omega_i\eta_i+\omega_j\eta_j))} (f C_iC_j)\cdot (B+J)(w)}{ \det \left(\frac{\partial \lambda^\alpha}{\partial w^b}\right)} \right)_{w = I_k} \nonumber
     \\&\sim \frac{1}{\la \kappa_i \kappa_j\ra}\times \frac{-\omega'}{4 \omega_i \omega_j} f^{AB}_C(C_i)_A(C_j)_B V^C(f')
\end{align}
The coefficient of the pole $1/\la \kappa_i \kappa_j \ra$ is recognisably $\frac{-\omega'}{4\omega_i \omega_j}$ times a vertex operator for $C'_A f'$, which is the wavefunction with external data
\begin{align}
    (\kappa')^\alpha &= \kappa_i^\alpha = \kappa_j^\alpha
    \\
    \omega' (\tilde \kappa')^{\dot \alpha} &= \omega_i\tilde\kappa_i^{\dot \alpha}+\omega_j\tilde\kappa_j^{\dot \alpha}
    \\
    \omega' \eta'_I &= \omega_i(\eta_i)_{I}+\omega_j(\eta_j)_I
    \\
    C'_A &= f_A^{BC}(C_i)_B (C_j)_C
\end{align}
Equivalently, we may instead express this system of equations in terms of the momentum and supermomentum (I have copied the color vector equation as well)
\begin{equation}
    P'^{\alpha \dot \alpha} = P_i^{\alpha \dot \alpha} + P_j^{\alpha \dot \alpha}, \quad \Pi'^{\alpha}_I = (\Pi_i)^{\alpha}_I + (\Pi_j)^{\alpha}_I, \quad C'_A = f_A^{BC}(C_i)_B (C_j)_C
\end{equation}
Stripping off the color vectors, we see that analysis of the collinear behaviour indeed recovers the expected tree-level splitting function of $\mathcal{N}=4$ super Yang-Mills
\begin{equation}
    \text{Split}(i,j) = \frac{f^{A_iB_j}_C}{\la \kappa_i \kappa_j\ra}\times \frac{-\omega'}{4 \omega_i \omega_j}
\end{equation}
as can be verified directly from the RSVW formula, up to a factor of $2$ that comes from wavefunction normalisation.
\pagebreak

\section{Discussion}\label{sec: Discussion}

We have constructed a free-field realisation of a current algebra on $\mathbb{CP}^1$ that generates only the single-trace terms in the evaluation of a conventional affine Kac-Moody current correlator. We investigated and explained how the use of the single-trace current algebra in the Berkovits-Witten action operationalises the removal of the Feynman diagrams with the exchange of conformal supergravity intermediates at tree level, allowing us to recover the pure $\mathcal{N}  = 4$ super Yang-Mills tree amplitudes. 

The construction of a BRST operator to quotient by a $\mathfrak{sl}(1|1)$ played a central role in the construction of the single-trace current system. Gauging $\mathfrak{sl}(1|1)$ currents in a twistor string has notable precedent. Firstly, the Mason-Skinner ambitwistor string \cite{Mason_2008} has a gauged $\mathfrak{sl}(1|1)$ in the gravity case. Secondly, a $\mathfrak{sl}(1|1)$ gauging\footnote{Brought to my attention by David Skinner} appears as a subalgebra of the gauged bilinears of $Y,Z,\rho^a$ in an analogous context in the Skinner $\mathcal{N}=8$ twistor string \cite{skinner2013twistorstringsn8supergravity}. There is no evidence that the repeated appearance of an $\mathfrak{sl}(1|1)$ in these twistor worldsheet settings is more than a coincidence, but it is intriguing and perhaps merits further examination.

The single-trace current system has immediate applications to $\mathbb{CP}^1$ models for 4d gluon scattering, such as in twistor worldsheet models at genus 0 or in Celestial Holography. For instance, use of the single-trace current system in place of a standard current algebra in the Mason-Skinner ambitwistor string \cite{Mason_2014} means that genus 0 string correlators reproduce tree-level Yang-Mills scattering precisely.

It is natural to ask if loop-level pure (super)Yang-Mills amplitudes can be computed by using the single-trace current system in the Berkovits-Witten twistor string and ambitwistor strings at higher genus. Preliminary results show that the $\mathfrak{psl}(1|1)$ and $\mathfrak{sl}(1|1)$ prescriptions no longer agree at higher genus: current correlators in the $\mathfrak{psl}(1|1)$ prescription appear to all vanish for $g > 0$, while the evaluation in the $\mathfrak{sl}(1|1)$ prescription is subtle (in part due to instantons of $a_J$) and we have no definitive statements at this point. It seems probable that to advance this programme beyond genus 0, there is a canonical choice of prescription with which to construct the single-trace current system, and the fact that the $\mathfrak{psl}(1|1)$ and $\mathfrak{sl}(1|1)$ prescriptions agreed at tree-level is a genus 0 coincidence.

An angle of attack that should shed light on the correct prescription is investigating the construction of the $\mathfrak{sl}(1|1)$ system in terms of brane insertions in the Witten twistor string \cite{Witten_2004}. Rather than introducing an \textit{ad-hoc} worldsheet current algebra as in the Berkovits-Witten twistor string, the free-fields (whose bilinears realise the affine Kac-Moody) of the Witten twistor string come from the worldvolume theory of a D1-brane holomorphically embedded in $\mathbb{PT}$. It would be tremendously interesting to understand if/how the single-trace current algebra arises from brane insertions in the Witten twistor string, and then also see how to import the resulting construction to the Berkovits-Witten twistor string.

Another instructive feature of the tree-level amplitude calculations from Berkovits-like twistor worldsheet models is that the key ingredient of the correlator evaluation is the localisation onto holomorphic maps from $\mathbb{CP}^1$ into curves on twistor space. With this in mind, at tree-level the only role of dynamical worldsheet gravity is to provide the $SL(2, \mathbb{C})$ quotient (encoding the redundant holomorphic reparameterisations of the worldsheet). Although this is desirable, including dynamical gravity forces us to introduce additional auxillary matter with which to cancel the central charge contributed by the ghosts for the stress tensor gauging, and it is unclear if the benefit outweighs this cost. Indulging in some numerology, in the application of the single-trace current system to the Berkovits-Witten twistor string we find that the central charge of the requisite auxiliary matter is precisely 26 - i.e, in the absence of the stress tensor gauging, \emph{no auxiliary matter} is required to cancel the central charge. It is therefore attractive to ask if we can take the conformal-gauge-fixed twistor string action as a starting point, as the definition of a chiral gauged $\beta\gamma$ system on $\mathbb{CP}^1$. Notable precedent for not gauging the stress tensor in the Berkovits-style twistor worldsheet models exists in the Skinner $\mathcal{N}=8$ twistor string, in which there is no integration over the bosonic worldsheet metric degrees of freedom.

Of course, the integrals over dynamical worldsheet gravity at higher genus are what provide a prescription for calculating higher-loop amplitudes. If we choose to work without dynamical worldsheet gravity and choose our starting point to be a gauged chiral $\beta\gamma$ system on $\mathbb{CP}^1$, we must furnish the theory with a mechanism with which to compute loop amplitudes from a genus 0 worldsheet, and there are no extant proposals for such a mechanism in Berkovits-style twistor strings.

It is worth mentioning, however that the \emph{nodal sphere} prescription of Geyer, Mason, Monteiro and Tourkine \cite{Geyer_2016,Geyer_2015,Adamo:2013tsa,Roehrig_2018} provides exactly such a mechanism for the \emph{ambitwistor} string. The result of this paper combined with the result of \cite{Roehrig_2018} gives a concrete proposal for a theory (a gauged chiral $\beta\gamma$ system) defined on $\mathbb{CP}^1$ whose correlators are 1-to-1 with scattering amplitudes of Yang-Mills at least at tree level and 1 loop. The action is the conformal-gauge-fixed 4d Mason-Skinner ambitwistor action \cite{Mason_2014} for Yang-Mills, with the single-trace current system $S_{\text{ff}}$, and the Roehrig-Skinner gluing operator \cite{Roehrig_2018} with Yang-Mills coupling constant $g_{\text{YM}}$
\begin{equation}\label{eq: 2d theory for yang mills}
    S = \int_{\mathbb{CP}^1} \left(P_\mu \bar \partial X^\mu + \eta_{\mu \nu} \psi^\mu \bar \partial \psi^\nu +e \frac{1}{2}P^2 + \chi P_\mu \psi^\mu \right)+ S_{\text{ff}} + g_{\text{YM}}^2 \int_{\mathbb{CP}^1 \times \mathbb{CP}^1} \Delta_{\text{YM}}(\sigma_1, \sigma_2)
\end{equation}
This is \textit{not} a string theory, but rather a chiral $\beta\gamma$ system defined on $\mathbb{CP}^1$, deformed by the highly nonlocal operator $\Delta_{\text{YM}}$ (defined in equation 4.1 of \cite{Roehrig_2018}). Using the standard integrated Mason-Skinner ambitwistor Yang-Mills vertex operators in this theory to compute correlators gives
\begin{equation}
    \la \text{Worldsheet correlator} \ra = \text{Vol}(SL(2, \mathbb{C}))\la \text{Scattering amplitude}\ra
\end{equation}
in perturbation theory at least up to one loop, but it is not known if the equality holds at higher loop order. At least in the case of tree-level MHV scattering, it can be shown (with some work) that the $\mathbb{CP}^1$ worldsheet can be identified with the Celestial sphere of Minkowski space. The extent to which this theory can be a satisfactory Celestial dual for 4d Yang-Mills is debatable, but it would be nonetheless interesting to investigate.

\begin{acknowledgments}
SS is supported by the Simons Collaboration on Celestial Holography and the Maxwell Institute of the University of Edinburgh. Many thanks to Tim Adamo, Atul Sharma, David Skinner,  Lionel Mason, Wei Bu, Iustin Surubaru, and Sonja Klisch for useful discussions and feedback on the draft. In particular, many thanks to Tim for many useful discussions and encouragement.
\end{acknowledgments}

\pagebreak
\appendix
\section{Field space symplectomorphism}\label{appendix: symplectomorphism}
The construction of $\mathfrak{psl}(1|1) \cong \mathbb{C}^{0|2}$ in subsubsection \ref{subsubsec: psl gaugings} did not come out of nowhere, but is a demonstration of a nice trick that can be done with these first order actions using the field theory analogue of canonical transformations in Lagrangian mechanics, i.e field space symplectomorphisms. Starting with the free action $S_{\text{ff}} +  \int u \bar \partial v$, a simple realisation of $\mathfrak{psl}(1|1) \cong\mathbb{C}^{0|2}$ is
\begin{equation}
    \Pi_0^+: u, \quad \Pi^-_0: \tilde \rho_a \phi^a, \quad \Pi_0^\pm \Pi_0^\pm\sim 0
\end{equation}
which can be consistently gauged and can be checked to obey the defining relations of $\mathfrak{psl}(1|1) \cong \mathbb{C}^{0|2}$. Gauging these currents, the action takes the form
\begin{equation}
    S_{\text{ff}} + \int u \bar \partial v + \int \chi_+u + \chi_-\tilde \rho_a \phi^a
\end{equation}
Now we will identify the $\mathbb{C}^{0|1}$ generated by $u$ on the auxiliary $u, v$ system with a $\mathbb{C}^{0|1}$ in the free field system generated by $\tilde \phi_a \rho^a$. Explicitly, consider the action of the Hamiltonian $v \tilde \phi_a \rho^a$ via the charge $Q_{\text{def}}:=\oint v \tilde \phi_a \rho^a$ on the Lagrangian, the currents, and the BRST-cohomology. The free Lagrangian from $S_{\text{ff}} + \int u \bar \partial v$ takes the form of the symplectic potential on field space and can be checked to be left undeformed by the charge.
\begin{equation}
    e^{Q_{\text{def}}} \left(\tilde \phi_a \bar \partial \phi^a + \tilde \rho_a \bar\partial \rho^a + u \bar \partial v\right)(z) = \left(\tilde \phi_a \bar \partial \phi^a + \tilde \rho_a \bar\partial \rho^a + u \bar \partial v\right)(z)
\end{equation}
The currents are more interesting. For $\Pi_0^+=u$, we have
\begin{align}
    Q_{\text{def}} \Pi_0^+ &=\left(\oint v \tilde \phi_a \rho^a\right) u = \tilde \phi_a \rho^a \implies Q_{\text{def}}^2 \Pi_0^+ = 0
     \\
    e^{Q_{\text{def}}}\Pi_0^+ &= \lim_{N' \rightarrow \infty}\left(1+\frac{1}{N'}Q_{\text{def}}\right)^{N'}u = u + \tilde \phi_a \rho^a = \Pi^+
\end{align}
Note that in this particular case, the evaluation of $e^{Q_{\text{def}}}$ was simple because $Q_{\text{def}}^2 \Pi_0^+ = 0$. In more general settings one would need to evaluate $(1+Q/N')^{N'}$ explicitly. 

For $\Pi_0^- = \tilde \rho_a \phi^a$, we have
\begin{align}
    Q_{\text{def}} \Pi_0^- &=\left(\oint v \tilde \phi_a \rho^a\right) \tilde \rho_a \phi^a = -v (\tilde\phi_a \phi^a + \tilde\rho_a \rho^a) +N \partial v \implies Q_{\text{def}}^2 \Pi_0^-=0
    \\
    e^{Q_{\text{def}}}\Pi_0^- &= \lim_{N' \rightarrow \infty}\left(1+\frac{1}{N'}Q_{\text{def}}\right)^{N'}\tilde \rho_a \phi^a = \tilde \rho_a \phi^a -v (\tilde\phi_a \phi^a + \tilde\rho_a \rho^a) +N \partial v = \Pi^-
\end{align}
The double pole with coefficient $\delta^a_a = N$ between $\tilde \phi_a \rho^a$ and $\tilde \rho_a \phi^a$ gives us the term $N \partial v$. In terms of the field theory, the action of the charge defines a field redefinition of the fundamental fields (which in this case has trivial Jacobian, which is implied by the nilpotence of $\delta_{Q_{def}}$ on all the fundamental fields) that preserves the free first order action; i.e, the field theory analogue of a canonical transformation. Since the transformation is therefore just a change of variables with trivial Jacobian that preserves the free action, the OPE relations between $\Pi^\pm_0$ in fact imply the OPE relations between $\Pi^\pm$. 

Studying the space of BRST-closed operators is also quite enlightening. Before taking the deformation, the BRST-closed fundamental fields and $\tilde \phi, \phi,\tilde\rho,\rho$ bilinears are given by $(\tilde \rho_a, \phi^b, u, \tilde \phi_a \phi^b + \tilde \rho_a \rho^b)$, while after the deformation, the BRST-closed combinations are
\begin{equation}
    \begin{pmatrix}
        \tilde \rho_a \\ \phi^b \\ u \\ \tilde \phi_a \phi^b + \tilde \rho_a \rho^b
    \end{pmatrix} \xrightarrow{e^{Q_{\text{def}}}} \begin{pmatrix}
        \tilde \rho_a - v \tilde \phi_a \\ \phi^b + v \rho^b\\ u + \tilde \phi_a \rho^a \\ \tilde \phi_a \phi^b + \tilde \rho_a \rho^b
    \end{pmatrix}.
\end{equation}
The first 2 lines are \textit{incidence relations}, and are the result of identifying the generators of $\mathbb{C}^{0|1}$ in its representation in terms of the action of $\tilde \phi_a \rho^a$ on the free fields and $u$ on the $v$. We call them incidence relations because the standard twistor incidence relations can be constructed in an identical way by identifying the action of translations $\subset$ the conformal group on $W_A=(\pi_{\dot \alpha}, \omega^\alpha),Z^A=(\mu^{\dot \alpha}, \lambda_\alpha)$ and on $p_{\alpha \dot \alpha}, x^{\alpha\dot \alpha}$ using the Hamiltonian $x^{\alpha \dot \alpha}\pi_{\dot \alpha}\lambda_\alpha$.
\section{The effective field theory (EFT) of the Berkovits-Witten twistor string}\label{appendix: BW twistor string}
This appendix accompanies subsection \ref{subsec: decoupling of csg} and provides the requisite background for understanding the EFT of the Berkovits-Witten twistor string, including the definitions and details that were omitted in the main text.

The correlators computed by the Berkovits-Witten twistor string (with a conventional Kac-Moody current algebra at level $k$) are consistent with tree amplitudes computed with the following twistor action (the precise form of the interaction term was first determined by Mason and Skinner \cite{Mason_2008} in equation 73 and the unnumbered equation that follows it, written here in very different notation).
\begin{multline}\label{eq: appendix EFT for BW}
    S_{BW}=\int \D^{3|4}Z \wedge \left(k \times hCS(\mathcal{A})-hCS(\partial_{\mathcal{M}} f^{\mathcal{N}}) + N(f) \lrcorner g\right) \\ + \int \d^{4|8}x \exp\left(\left(\int_{\mathcal{L}_x} g \right)+ k\log \det(\bar \partial + \mathcal{A}) - \log \det(\bar \partial + {\partial_{\mathcal{M}} f^{\mathcal{N}}})\right),
\end{multline}
\begin{align}
    f &:= f^{\mathcal{M}} \partial_{\mathcal{M}} \in \Omega^{0,1}(\mathbb{PT}, T^{1,0}\mathbb{PT}), \quad \partial_{\mathcal{M}}f^{\mathcal{M}} = 0, \quad N(f) := (\bar \partial f^\mathcal{M} + \mathcal{L}_f f^\mathcal{M})\partial_\mathcal{M}
    \\
    g &:= g_\mathcal{M} \d Z^{\mathcal{M}} \in \Omega^{1,1}(\mathbb{PT})
    \\
    \mathcal{A} & \in \Omega^{0,1}(\mathbb{PT},\mathfrak{g}), \quad \mathfrak{g} \subset \mathfrak{sl}(N), \quad \quad \partial_{\mathcal{M}} f^\mathcal{N} \in \Omega^{0,1}(\mathbb{PT}, \mathfrak{sl}(4|4))
\end{align}
in which $hCS(\mathcal{A})$ is the holomorphic-Chern-Simons 3-form with respect to the $\bar D:= \bar \partial + \mathcal{L}_{f}$ operator
\begin{equation}
    hCS(\mathcal{A}):=\int \D^{3|4}Z \wedge \text{tr}\left(\mathcal{A} \bar D \mathcal{A} + \frac{2}{3}\mathcal{A}^3\right), \quad \quad \bar D:= \bar \partial + \mathcal{L}_{f \cdot \partial}
\end{equation}
The term $hCS(\partial_\mathcal{M}f^\mathcal{N})$ means to treat $\partial_\mathcal{M}f^\mathcal{N}$ as a partial connection valued in $\mathfrak{sl}(4|4)$ and plug it into the $hCS$ functional. Due to the divergence-free constraint $\partial_\mathcal{M}f^\mathcal{M}=0$, the quadratic term of $hCS(\partial_\mathcal{M}f^\mathcal{N})$ is in fact a total holomorphic derivative, and can be neglected.

The $\log \det$ of a partial connection $\Omega^{0,1}(\mathbb{PT}, \mathfrak{g})$ is defined to be
\begin{equation}
    \log \det(\bar \partial + \mathcal{A}) := \int_{\mathcal{L}_x \times \mathcal{L}_x} \frac{\D\sigma_1\D\sigma_2}{\la \sigma_1 \sigma_2\rangle^2} \text{tr}\left(H\mathcal{A}H^{-1}(Z(\sigma_1))H\mathcal{A}H^{-1}(Z(\sigma_2))\right), \quad \bar \partial_{\mathcal{L}_x} H = \mathcal{A}|_{\mathcal{L}_x}H. 
\end{equation}
 In which $\mathcal{L}_x$ is the space of holomorphic curves with respect to $\bar D:=\bar \partial + \mathcal{L}_f$, as a small deformation away from the flat twistor lines $L_x$
\begin{multline}
    L_x: \left\{\begin{pmatrix}
        \mu^{\dot \alpha} \\\lambda_\alpha \\ \psi^I
    \end{pmatrix} -\begin{pmatrix}
        x^{\dot \alpha \alpha}\lambda_\alpha \\ \lambda_\alpha \\ \theta^{I \alpha} \lambda_\alpha
    \end{pmatrix} = 0\right\}, \xrightarrow{\text{deform}}
    \\
    \mathcal{L}_x: \left\{\begin{pmatrix}
        \mu^{\dot \alpha} \\\lambda_\alpha \\ \psi^I
    \end{pmatrix} -\begin{pmatrix}
        x^{\dot \alpha \alpha}\lambda_\alpha \\ \lambda_\alpha \\ \theta^{I \alpha} \lambda_\alpha
    \end{pmatrix} - \underbrace{\int_{\mathcal{L}_x} \frac{\D \lambda'}{\la \lambda \lambda'\ra}\frac{\la \lambda a \ra\la \lambda b \ra}{\la \lambda' a \ra\la \lambda' b \ra} f^\mathcal{M}(x,\theta,\lambda')}_{\frac{1}{\bar D} f^\mathcal{M}:=} = 0\right\}
\end{multline}
Although the definition for $\mathcal{L}_x$ is recursive, this will suffice for amplitude calculations in perturbation theory. Expanding an integrand on a curved twistor line $\int_{\mathcal{L}_x } I$ in orders of $f^\mathcal{M}$ around $\int_{L_x } I$, we have
\begin{multline}
    \int_{\mathcal{L}_x } I = \int_{L_x } \left(I + \left(\frac{1}{\bar D} f^\mathcal{M}\right) \left(\frac{\partial I}{\partial Z^\mathcal{M}}\right)\right) +\mathcal{O}(f^2I)
    \\
    =\int_{L_x } \left(I + \left(\int_{L_x} \frac{\D \lambda'}{\la \lambda \lambda'\ra}\frac{\la \lambda a \ra\la \lambda b \ra}{\la \lambda' a \ra\la \lambda' b \ra} f^\mathcal{M}(x,\theta,\lambda')\right) \left(\frac{\partial I}{\partial Z^\mathcal{M}}\right)\right) +\mathcal{O}(f^2I)
\end{multline}
where in going to the second line we have expanded the $\mathcal{L}_x$ in the definition of $\frac{1}{\bar \D} f^{\mathcal{M}}$ and kept the zeroth order term. From the worldsheet theory point of view, these terms arise from the free field OPEs between the vertex operator for $f$ and other vertex operators $V(Z(\sigma))$ that are functions only of $Z$
\begin{equation}
    V_f := \int f^\mathcal{M}(Z(\sigma)) W_\mathcal{M}(\sigma), \quad W_{\mathcal{M}}(\sigma_1) V(Z(\sigma_2)) \sim \frac{\D \sigma_1}{(\sigma_1\sigma_2)}\frac{(\sigma_2 a)}{(\sigma_1 a)} \frac{\la\lambda(\sigma_2) b \ra}{\la\lambda(\sigma_1) b \ra} \frac{\partial V(Z(\sigma_2))}{\partial Z^{\mathcal{M}}(\sigma_2)}
\end{equation}
We have used the $W_{\mathcal{M}}\in K_{\mathbb{CP}^1}\times \mathcal{L}^{-1}_d$, $Z^{\mathcal{N}}\in \mathcal{L}_d$ OPE, which is
\begin{equation}
    W_{\mathcal{M}}(\sigma_1)Z^\mathcal{N}(\sigma_2) \sim \delta_{\mathcal{M}}^{\,\,\mathcal{N}} \underbrace{\frac{\sqrt{\D \sigma_1}\sqrt{\D \sigma_2}}{(\sigma_1\sigma_2)}}_{\text{OPE for }K^{1/2}K^{1/2} }\underbrace{\frac{\sqrt{\D \sigma_1}(\sigma_2 a)}{\sqrt{\D \sigma_2}(\sigma_1 a)}}_{\text{twist by } K^{1/2}} \underbrace{\frac{\la\lambda(\sigma_2) b \ra}{\la\lambda(\sigma_1) b \ra}}_{\text{twist by }\mathcal{L}^{-1}_d}
\end{equation}
with arbitrary constant reference spinors $a^a, b^\alpha$ on a patch of the Riemann sphere in which $(\sigma a) \neq0$ and $\la \lambda(\sigma) b \ra \neq 0$, and $\delta_\mathcal{M}^{\,\,\mathcal{N}}=\text{diag}(-\delta_m^{\,\,n}, \delta_i^{\,\,j})$.
\pagebreak
\bibliography{bibliography}

\providecommand{\href}[2]{#2}\begingroup\raggedright\begin{thebibliography}{10}

\bibitem{Parke:1986gb}
S.J.~Parke and T.R.~Taylor, \emph{{An Amplitude for $n$ Gluon Scattering}}, \href{https://doi.org/10.1103/PhysRevLett.56.2459}{\emph{Phys. Rev. Lett.} {\bfseries 56} (1986) 2459}.

\bibitem{Berends:1987me}
F.A.~Berends and W.T.~Giele, \emph{{Recursive Calculations for Processes with n Gluons}}, \href{https://doi.org/10.1016/0550-3213(88)90442-7}{\emph{Nucl. Phys. B} {\bfseries 306} (1988) 759}.

\bibitem{Nair:1988bq}
V.P.~Nair, \emph{{A Current Algebra for Some Gauge Theory Amplitudes}}, \href{https://doi.org/10.1016/0370-2693(88)91471-2}{\emph{Phys. Lett. B} {\bfseries 214} (1988) 215}.

\bibitem{Witten_2004}
\emph{Perturbative gauge theory as a string theory in twistor space}, \href{https://doi.org/10.1007/s00220-004-1187-3}{\emph{Commun. Math. Phys.} {\bfseries 252} (2004) 189} [\href{https://arxiv.org/abs/hep-th/0312171}{{\ttfamily hep-th/0312171}}].

\bibitem{Berkovits_2004}
N.~Berkovits, \emph{{An Alternative string theory in twistor space for N=4 superYang-Mills}}, \href{https://doi.org/10.1103/PhysRevLett.93.011601}{\emph{Phys. Rev. Lett.} {\bfseries 93} (2004) 011601} [\href{https://arxiv.org/abs/hep-th/0402045}{{\ttfamily hep-th/0402045}}].

\bibitem{Berkovits_2004confsugra}
N.~Berkovits and E.~Witten, \emph{{Conformal supergravity in twistor-string theory}}, \href{https://doi.org/10.1088/1126-6708/2004/08/009}{\emph{JHEP} {\bfseries 08} (2004) 009} [\href{https://arxiv.org/abs/hep-th/0406051}{{\ttfamily hep-th/0406051}}].

\bibitem{Roiban_2004}
R.~Roiban, M.~Spradlin and A.~Volovich, \emph{{On the tree level S matrix of Yang-Mills theory}}, \href{https://doi.org/10.1103/PhysRevD.70.026009}{\emph{Phys. Rev. D} {\bfseries 70} (2004) 026009} [\href{https://arxiv.org/abs/hep-th/0403190}{{\ttfamily hep-th/0403190}}].

\bibitem{Mason_2014}
L.~Mason and D.~Skinner, \emph{{Ambitwistor strings and the scattering equations}}, \href{https://doi.org/10.1007/JHEP07(2014)048}{\emph{JHEP} {\bfseries 07} (2014) 048} [\href{https://arxiv.org/abs/1311.2564}{{\ttfamily 1311.2564}}].

\bibitem{Berkovits_2014}
N.~Berkovits, \emph{{Infinite Tension Limit of the Pure Spinor Superstring}}, \href{https://doi.org/10.1007/JHEP03(2014)017}{\emph{JHEP} {\bfseries 03} (2014) 017} [\href{https://arxiv.org/abs/1311.4156}{{\ttfamily 1311.4156}}].

\bibitem{siegel2004untwistingtwistorsuperstring}
W.~Siegel, \emph{{Untwisting the twistor superstring}},  \href{https://arxiv.org/abs/0404255}{{\ttfamily 0404255}}.

\bibitem{Adamo:2014wea}
T.~Adamo, E.~Casali and D.~Skinner, \emph{{A Worldsheet Theory for Supergravity}}, \href{https://doi.org/10.1007/JHEP02(2015)116}{\emph{JHEP} {\bfseries 02} (2015) 116} [\href{https://arxiv.org/abs/1409.5656}{{\ttfamily 1409.5656}}].

\bibitem{Magnea_2021}
L.~Magnea, \emph{{Non-abelian infrared divergences on the celestial sphere}}, \href{https://doi.org/10.1007/JHEP05(2021)282}{\emph{JHEP} {\bfseries 05} (2021) 282} [\href{https://arxiv.org/abs/2104.10254}{{\ttfamily 2104.10254}}].

\bibitem{Bu:2023vjt}
W.~Bu and S.~Seet, \emph{{A hidden 2d CFT for self-dual Yang-Mills on the celestial sphere}}, \href{https://doi.org/10.1007/JHEP08(2024)022}{\emph{JHEP} {\bfseries 08} (2024) 022} [\href{https://arxiv.org/abs/2310.17457}{{\ttfamily 2310.17457}}].

\bibitem{Melton:2024akx}
W.~Melton, A.~Sharma, A.~Strominger and T.~Wang, \emph{{Celestial Dual for Maximal Helicity Violating Amplitudes}}, \href{https://doi.org/10.1103/PhysRevLett.133.091603}{\emph{Phys. Rev. Lett.} {\bfseries 133} (2024) 091603} [\href{https://arxiv.org/abs/2403.18896}{{\ttfamily 2403.18896}}].

\bibitem{Seet:2024vmh}
S.~Seet, \emph{{Twistor Space and Celestial Holography}}, Ph.D. thesis, Cambridge U., DAMTP, 2024.
\newblock 10.17863/CAM.112793.

\bibitem{Geyer:2015bja}
Y.~Geyer, L.~Mason, R.~Monteiro and P.~Tourkine, \emph{{Loop Integrands for Scattering Amplitudes from the Riemann Sphere}}, \href{https://doi.org/10.1103/PhysRevLett.115.121603}{\emph{Phys. Rev. Lett.} {\bfseries 115} (2015) 121603} [\href{https://arxiv.org/abs/1507.00321}{{\ttfamily 1507.00321}}].

\bibitem{Roehrig_2018}
K.A.~Roehrig and D.~Skinner, \emph{{A Gluing Operator for the Ambitwistor String}}, \href{https://doi.org/10.1007/JHEP01(2018)069}{\emph{JHEP} {\bfseries 01} (2018) 069} [\href{https://arxiv.org/abs/1709.03262}{{\ttfamily 1709.03262}}].

\bibitem{Mason_2008}
L.J.~Mason and D.~Skinner, \emph{{Heterotic twistor-string theory}}, \href{https://doi.org/10.1016/j.nuclphysb.2007.11.010}{\emph{Nucl. Phys. B} {\bfseries 795} (2008) 105} [\href{https://arxiv.org/abs/0708.2276}{{\ttfamily 0708.2276}}].

\bibitem{Wakimoto:1986gf}
M.~Wakimoto, \emph{{Fock representations of the affine lie algebra A1(1)}}, \href{https://doi.org/10.1007/BF01211068}{\emph{Commun. Math. Phys.} {\bfseries 104} (1986) 605}.

\bibitem{FRIEDAN198693}
D.~Friedan, E.~Martinec and S.~Shenker, \emph{Conformal invariance, supersymmetry and string theory}, \href{https://doi.org/https://doi.org/10.1016/S0550-3213(86)80006-2}{\emph{Nuclear Physics B} {\bfseries 271} (1986) 93}.

\bibitem{VERLINDE198795}
E.~Verlinde and H.~Verlinde, \emph{Multiloop calculations in covariant superstring theory}, \href{https://doi.org/https://doi.org/10.1016/0370-2693(87)91148-8}{\emph{Physics Letters B} {\bfseries 192} (1987) 95}.

\bibitem{witten2023superstringperturbationtheoryrevisited}
E.~Witten, \emph{{Superstring Perturbation Theory Revisited}},  \href{https://arxiv.org/abs/1209.5461}{{\ttfamily 1209.5461}}.

\bibitem{ZEILBERGER198561}
D.~Zeilberger, \emph{A combinatorial approach to matrix algebra}, \href{https://doi.org/https://doi.org/10.1016/0012-365X(85)90192-X}{\emph{Discrete Mathematics} {\bfseries 56} (1985) 61}.

\bibitem{Reid-Edwards:2012vwx}
R.A.~Reid-Edwards, \emph{{On Closed Twistor String Theory}},  \href{https://arxiv.org/abs/1212.6047}{{\ttfamily 1212.6047}}.

\bibitem{Adamo:2021zpw}
T.~Adamo, W.~Bu, E.~Casali and A.~Sharma, \emph{{Celestial operator products from the worldsheet}}, \href{https://doi.org/10.1007/JHEP06(2022)052}{\emph{JHEP} {\bfseries 06} (2022) 052} [\href{https://arxiv.org/abs/2111.02279}{{\ttfamily 2111.02279}}].

\bibitem{Bu:2021avc}
W.~Bu, \emph{{Supersymmetric celestial OPEs and soft algebras from the ambitwistor string worldsheet}}, \href{https://doi.org/10.1103/PhysRevD.105.126029}{\emph{Phys. Rev. D} {\bfseries 105} (2022) 126029} [\href{https://arxiv.org/abs/2111.15584}{{\ttfamily 2111.15584}}].

\bibitem{skinner2013twistorstringsn8supergravity}
D.~Skinner, \emph{{Twistor strings for $ \mathcal{N} $ = 8 supergravity}}, \href{https://doi.org/10.1007/JHEP04(2020)047}{\emph{JHEP} {\bfseries 04} (2020) 047} [\href{https://arxiv.org/abs/1301.0868}{{\ttfamily 1301.0868}}].

\bibitem{Geyer_2016}
Y.~Geyer, L.~Mason, R.~Monteiro and P.~Tourkine, \emph{{One-loop amplitudes on the Riemann sphere}}, \href{https://doi.org/10.1007/JHEP03(2016)114}{\emph{JHEP} {\bfseries 03} (2016) 114} [\href{https://arxiv.org/abs/1511.06315}{{\ttfamily 1511.06315}}].

\bibitem{Geyer_2015}
Y.~Geyer, L.~Mason, R.~Monteiro and P.~Tourkine, \emph{{Loop Integrands for Scattering Amplitudes from the Riemann Sphere}}, \href{https://doi.org/10.1103/PhysRevLett.115.121603}{\emph{Phys. Rev. Lett.} {\bfseries 115} (2015) 121603} [\href{https://arxiv.org/abs/1507.00321}{{\ttfamily 1507.00321}}].

\bibitem{Adamo:2013tsa}
T.~Adamo, E.~Casali and D.~Skinner, \emph{{Ambitwistor strings and the scattering equations at one loop}}, \href{https://doi.org/10.1007/JHEP04(2014)104}{\emph{JHEP} {\bfseries 04} (2014) 104} [\href{https://arxiv.org/abs/1312.3828}{{\ttfamily 1312.3828}}].

\end{thebibliography}\endgroup
\end{document}